\documentclass{emulateapj}
\usepackage{lscape}
\usepackage{natbib}

\def\OII{[{\ion{O}{2}}]}

\def\OIHa{[{\ion{O}{1}}]/H$\alpha$}
\def\NIIHa{[{\ion{N}{2}}]/H$\alpha$}
\def\SIIHa{[{\ion{S}{2}}]/H$\alpha$}

\def\OIIIHb{[{\ion{O}{3}}]/H$\beta$}
\def\OIII5007Hb{[{\ion{O}{3}}] $\lambda5007$/H$\beta$}
\def\4959_5007{[\ion{O}{3}] $\lambda \lambda$4959,5007}
\def\OIII49595007{[\ion{O}{3}] $\lambda \lambda 4959,5007$}
\def\ratioR23{([\ion{O}{2}] $\lambda$3727 +[\ion{O}{3}] $\lambda\lambda$4959,5007)/H$\beta$}
\def\R23{${\rm R}_{23}$}
\def\dS23{${\rm S}_{23}$}
\def\OIIIl{[\ion{O}{3}] $\lambda$5007}

\def\Msun{${\rm M}_{\odot}$}

\def\NII{[{\ion{N}{2}}]}

\def\NIIOII{[\ion{N}{2}]/[\ion{O}{2}]}
\def\ratioNIIOII{[\ion{N}{2}] $\lambda$6584/[\ion{O}{2}] $\lambda$3727}
\def\ratioNIIHa{[\ion{N}{2}] $\lambda$6584/H$\alpha$}
\def\OH{$12+\log({\rm O/H})$}

\def\ratioS23{([\ion{S}{2}] $\lambda \lambda$6717,31 +[\ion{S}{3}] $\lambda\lambda$9069,9532)/H$\beta$}
\def\NIIHa{[\ion{N}{2}]/H$\alpha$}

\def\SII{[{\ion{S}{2}}]}
\def\SIII{[{\ion{S}{3}}]}
\def\Hb{{H$\beta$}}
\def\O4363{[{\ion{O}{3}}] $\lambda$4363}
\def\OIII{[{\ion{O}{3}}]}

\def\Ha{{H$\alpha$}}
\def\L60{L$_{60}$}

\def\HII{\ion{H}{2}}
\def\B26{${{\rm B}_{26}}$}
\def\I26{${\rm I_{B_{26}}}$}

\def\Te{T$_{e}$}

\shorttitle{}
\shortauthors{}

\begin{document}

\title{Metallicity Calibrations and the Mass-Metallicity Relation for Star-Forming Galaxies}

\author{Lisa J. Kewley\altaffilmark{1}}
\affil {Institute of Astronomy, University of Hawaii}
\email {kewley@ifa.hawaii.edu}
\altaffiltext{1}{Hubble Fellow}

\author{Sara L. Ellison}
\affil{University of Victoria}

\begin{abstract}
We investigate the effect of metallicity calibrations,  AGN classification, and aperture covering fraction on the local mass-metallicity relation using 27,730 star-forming galaxies from the Sloan Digital Sky Survey (SDSS) Data Release 4.    We analyse the SDSS mass-metallicity relation with 10 metallicity calibrations, including theoretical and empirical methods.   We show that the choice of metallicity calibration has a significant effect on the shape and y-intercept(\OH) of the mass-metallicity relation.  The absolute metallicity scale (y-intercept) varies up to 
$\Delta[\log({\rm O/H})]= 0.7$~dex, depending on the calibration used, and the change in shape is substantial.  These results indicate that it is critical to use the same metallicity calibration when comparing different luminosity-metallicity or mass-metallicity relations.    We present new metallicity conversions that allow metallicities that have been derived using different strong-line calibrations to be converted to the same base calibration.  These conversions facilitate comparisons between different samples, particularly comparisons between galaxies at different redshifts for which different suites of emission-lines are available.   Our new conversions successfully remove the large $0.7$~dex discrepancies between the metallicity calibrations, and we reach agreement in the mass-metallicity relation to within $0.03$~dex on average.      We investigate the effect of AGN classification and aperture covering fraction on the mass-metallicity relation.  We find that different AGN classification methods have negligible effect on the SDSS MZ-relation.  We compare the SDSS mass-metallicity relation with nuclear and global relations from the Nearby Field Galaxy Survey (NFGS).   The turn over of the mass-metallicity relation at ${\rm M}_{*}\sim 10^{10}$~\Msun\ depends on aperture covering fraction.  We find that a lower redshift limit of $z<0.04$ is insufficient for avoiding aperture effects in fiber spectra of the highest stellar mass (${\rm M}_{*} > 10^{10}$~\Msun) galaxies.
\end{abstract}


\section{Introduction}
The relationship between metallicity and stellar mass provides crucial insight into galaxy formation and evolution.   Theory predicts that as time progresses, the mean stellar metallicity of galaxies increases with age as galaxies undergo chemical enrichment, while  the stellar mass of a galaxy will increase with time as galaxies are built through merging processes \citep[e.g.,][and references therein]{Pei95,Somerville99,Somerville00,Nagamine01,Calura04}.    A correlation between mass and metallicity arises if low mass galaxies have larger gas fractions than higher mass galaxies, as is observed in local galaxies \citep{McGaugh97,Bell00,Boselli01}.  The detailed relationship between metallicity and mass may depend critically on galactic-scale outflows driven by supernovae and stellar winds  \citep[see e.g.,][for a review]{Garnett02,Pettini02}.  Thus, robust measurements of the mass-metallicity (MZ) relation may provide important clues into the impact of galactic-scale winds on the chemical history of galaxies.

The MZ relation was first observed in irregular and blue compact galaxies \citep{Lequeux79,Kinman81}.  In subsequent work, luminosity was often used as a surrogate for mass because obtaining reliable mass estimates for galaxies was non-trivial.   \citet{Rubin84} provided the first evidence that metallicity is correlated with luminosity in disk galaxies.  Further investigations solidified the correlation between luminosity and metallicity in nearby disk galaxies \citep{Bothun84,Wyse85, Skillman89,Vila92, Zaritsky94,Garnett02}.    However, optical luminosity may not be a reliable surrogate for the stellar mass of a galaxy because optical luminosities are sensitive to the level of current star formation and are extinguished by dust.  Near infrared luminosities can be influenced by the age of the stellar population of a galaxy.   Fortunately, reliable stellar mass estimates are now possible, thanks to new state-of-the-art stellar evolutionary synthesis models \citep[e.g.,][]{Silva98,Leitherer99,Fioc99,Bruzual03}.  

Key insight into the mass-metallicity relation has recently been obtained with large spectroscopic surveys such as the Sloan Digital Sky Survey (SDSS) and the  2 degree Field Galaxy Redshift Survey (2dFGRS) \citep[e.g.,][]{Baldry02,Schulte03,Lamareille04,Tremonti04,Gallazzi05}.  Using the SDSS stellar masses, \citet[][; hereafter T04]{Tremonti04} characterized the local MZ relation for $\sim 53,000$ local galaxies.  The MZ relation is steep for masses  $\lesssim 10^{10.5}$~\Msun\ and flattens at higher masses.  T04 use chemical evolution models to interpret this flattening in terms of efficient galactic scale winds that remove metals from low mass galaxies ($M \lesssim 10^{10.5}$~\Msun).   Hierarchical galaxy formation models that include chemical evolution and feedback processes can reproduce the observed MZ relation \citep{DeLucia04,DeRossi06,Finlator07}.  However, these models rely on free parameters, such as feedback efficiency, that are relatively unconstrained by observations.   Alternative scenarios proposed to explain the MZ relation include low star formation efficiencies in low-mass galaxies caused by supernova feedback \citep{Brooks07}, and a variable integrated stellar initial mass function \citep{Koppen07}.   

The advent of large 8-10~m telescopes and efficient multi-object spectrographs enables the luminosity-metallicity (LZ) and, in some cases, the mass-metallicity relation to be characterized to high redshifts \citep{Kobulnicky99a,Carollo01,Pettini01,Lilly03,Kobulnicky03,Kobulnicky04,Shapley04,Maier04,Liang04,Maier05,Hoyos05,Savaglio05,Mouhcine06,Erb06,Maier06,Liang06}.   Evolution in the LZ and MZ relations are now predicted
by semi-analytic models of galaxy formation within the $\Lambda$ cold dark matter framework that include chemical hydrodynamic simulations   \citep{DeLucia04,Tissera05,DeRossi06,Dave07}.   Therefore reliable observational estimates of the LZ and MZ relations may provide important constraints on galaxy evolution theory.

Reliable MZ relations require a robust metallicity calibration.  Common metallicity calibrations are based on metallicity-sensitive optical emission-line ratios.  These calibrations include theoretical methods based on photoionization models \citep[see e.g.,][for a review]{Kewley02a}, empirical methods based on measurements of the electron-temperature of the gas,  \citep[e.g.,][]{Pilyugin01,Pettini04}, or a combination of the two \citep[e.g.,][]{Denicolo02}.  Comparisons among the metallicities estimated using these methods reveal large discrepancies \citep[e.g.,][]{Pilyugin01,Bresolin04, Garnett04}.  These discrepancies usually manifest as a systematic offset in metallicity estimates, with high values estimated by theoretical calibrations and lower metallicities estimated by electron-temperature metallicities.  Such offsets are found to be as large as 0.6~dex in log(O/H) units \citep{Liang06,Yin07} and may significantly affect the shape and zero-point of the mass-metallicity or luminosity-metallicity relations. 

Initial investigations into the extent of these discrepancies have recently been made by \citet{Liang06,Yin07,Yin07b}, and \citet{Nagao06}.    \citet{Liang06} applied the metallicity calibrations from four authors to $\sim40,000$ galaxies from the SDSS.  They showed that calibrations based on electron temperature metallicities produce discrepant metallicities when compared with calibrations based on photoionization models.   \citet{Yin07} compared the theoretical metallicities derived by T04 with \Te-based metallicities from \citet{Pilyugin01} and \citet{Pilyugin05}. They found a discrepancy of $\Delta[\log({\rm O/H})]=0.2$~dex between the two \Te-based metallicities, and a larger discrepancy of $\Delta[\log({\rm O/H})]=0.6$~dex between the \Te-based methods and the
theoretical method.     Similar results were obtained by  \citet{Yin07b} who extended the \citet{Liang06} work to low metallicities typical of low mass galaxies.

The cause of the metallicity calibration discrepancies remains unclear.   The discrepancy has been attributed to either an unknown problem with the photoionization models \citep{Kennicutt03}, or to temperature gradients or fluctuations that may cause metallicities based on the electron temperature method to underestimate the true metallicities \citep{Stasinska02,Stasinska05,Bresolin06}.  Until this discrepancy is resolved, the absolute metallicity scale is uncertain.

The metallicity discrepancy issue highlights the need for an in-depth study into the effect of metallicity calibration discrepancies and other effects on the MZ relation.   In this paper, we investigate the robustness of the local MZ relation for star-forming galaxies.  We focus on three important factors that may influence the shape, y-intercept, and scatter of the mass-metallicity relation:  choice of metallicity calibration, aperture covering fraction, and AGN removal method.   Our sample selection is described in Section~\ref{Sample}.  We describe the stellar mass estimates and metallicities in Sections~\ref{stellar_masses} and ~\ref{Metallicity}, respectively.
  We compare the mass-metallicity relation derived using 10 different, popular metallicity calibrations in Section~\ref{calibrations}.    We find larger discrepancies between the MZ relations derived with different metallicity calibrations than previously found.  We calibrate the discrepancies between the different calibrations using robust fits, and we provide conversion relations for removing these discrepancies in Section~\ref{Conversions}. We show that our new conversions  successfully remove the large metallicity discrepancies in the MZ relation.  We investigate the effect of different schemes for AGN removal in Section~\ref{AGN}, and we determine the effect of fiber covering fraction in Section~\ref{aperture}.  We discuss the impact of our results on the mass-metallicity and luminosity-metallicity relations in Section~\ref{Discussion}.  Our conclusions are given in Section~\ref{Conclusions}.   In the Appendix, we provide detailed descriptions of the metallicity calibrations used in this study and worked examples of the application of our conversions.   Throughout this paper, we adopt the flat $\Lambda$-dominated cosmology as measured by the WMAP experiment \citep[$h=0.72$, $\Omega_{m}=0.29$;][]{Spergel03}).


\section{Sample Selection}\label{Sample}

We selected our sample from the SDSS Data Release 4 (DR4) according to the following criteria:

\begin{enumerate}
\item Signal-to-noise (S/N) ratio of at least 8 in the strong emission-lines \OII~$\lambda \lambda3726,9$, \Hb, \OIII~$\lambda 5007$, \Ha, \NII~$\lambda 6584$, and \SII~$\lambda \lambda 6717,31$.  A S/N$>8$ is required for reliable metallicity estimates using established metallicity calibrations \citep{Kobulnicky99a}.   For each line, we define the S/N as the ratio of the statistical error on the flux to the total flux, where the statistical errors are calculated by the SDSS pipeline described in \citet{Tremonti04}.
\item Fiber covering fraction $> 20$\% of the total photometric g'-band light.  We use the raw DR4 fiber and Petrosian magnitudes to calculate the fiber covering fraction.   \citet{Kewley05a} found that a flux covering fraction $>20$\% is required for metallicities to begin to approximate global values.   Lower covering fractions can produce significant discrepancies between fixed-sized aperture and global metallicity estimates.   A covering fraction of $> 20$\% corresponds to a lower redshift limit of $z> 0.04$ for normal star-forming galaxies observed through the 3" SDSS fibers.   Large, luminous star-forming 
galaxies larger redshifts to satisfy the covering fraction $> 20$\% requiremend.  We investigate residual aperture effects in Section~\ref{aperture}.
\item Upper redshift limit $z<0.1$.  The SDSS star-forming sample becomes incomplete at 
redshifts above $z>0.1$ \citep[see e.g.,][]{Kewley06}.  With this upper redshift limit, the median redshift of our sample is $z\sim 0.068$.  
\item Stellar mass estimates must be available.  Stellar masses were derived by \citet{Kauffmann03a} and \citet{Tremonti04}.
\end{enumerate}

We remove galaxies containing AGN from our sample using the optical classification criteria given 
in \citet{Kewley06}.  This classification scheme utilizes optical strong line ratios to segregate 
galaxies containing AGN from galaxies dominated by star-formation.  A total of 84\% of the SDSS sample satisfying our selection criteria are star-forming according.  This fraction of star-forming galaxies differs from the fraction in \citet{Kewley06} because we apply a more stringent S/N cut which removes many LINERs prior to classification.   In Section~\ref{AGN}, we investigate different AGN classification schemes and their effect on the shape of the MZ relation.

The resulting sample contains 27,730 star-forming galaxies and does not include duplicates found in the original DR4 catalog.    We note that our sample is smaller than the \citet{Tremonti04} sample because we apply a more stringent S/N criterion and a stricter redshift range (T04 apply a redshift range of $0.005 < z < 0.25$).

We use the publically available emission-line fluxes that were calculated by the MPA/JHU group. \citep[described in][]{Tremonti04}.  These emission-line fluxes were calculated using a sophistcated code that is optimized for use with the SDSS galaxy spectra.  This code applies a least-squares fit of the \citet{Bruzual03} stellar population synthesis models and dust attenuation to the stellar continuum.  Once the continuum has been removed, the emission-line fluxes are fit with Gaussians, constraining the width and velocity separation of the Balmer lines together, and similarly for the forbidden lines.  

We correct the emission-line fluxes for extinction using the Balmer decrement and the \citet{Cardelli89} reddening curve.  We assume an ${\rm R_{V}=Av/{\rm E}(B-V)} = 3.1$ and an 
intrinsic H$\alpha$/H$\beta$ ratio of 2.85 \citep[the Balmer decrement for case B 
recombination at T$=10^4$K and $n_{e} \sim 10^2 - 10^4 {\rm cm}^{-3}$;][]{Osterbrock89}.  A total of 539 (2\%) of galaxies in our sample have Balmer decrements less than the theoretical value.  A Balmer decrement less than the theoretical value can result from an intrinsically low reddening combined with errors in the stellar absorption correction and/or errors in the line flux calibration and measurement.  For the S/N of our data, the lowest E(B-V) measurable is 0.01.  We therefore
assign these 539 galaxies an upper limit of E(B-V)$<0.01$.  

To investigate aperture effects (Section~\ref{aperture}), we compare the SDSS mass-metallicity relation
with the mass-metallicity relation derived from the Nearby Field Galaxy Survey (NFGS) \citep{Jansen00b,Jansen00a}.   
\citet{Jansen00a} selected the NFGS objectively from the CfA1 redshift survey 
\citep{Davis83,Huchra83} to approximate the local galaxy 
luminosity function \citep[e.g.,][]{Marzke94}.  The 198-galaxy NFGS sample contains the full range in Hubble type and absolute magnitude present in the CfA1 galaxy survey.

\citet{Jansen00b} provide integrated and nuclear spectrophotometry 
for almost all galaxies in the NFGS sample. 
The covering fraction and metallicities of the nuclear and global spectra for the NFGS are described in 
\citet{Kewley05a}.  The nuclear $B_{26}$  covering fraction ranges between 0.4 - 72\%, with an average nuclear covering fraction of $10\pm 11$\%\footnote[1]{The error quoted on the covering fraction is the standard error of the mean}.  The covering fraction of the integrated (global) spectra is between 52-97\% of the B-band light, with an average of is 82$\pm 7$\%.   

We apply the same extinction correction and AGN removal scheme to our NFGS supplementary sample as applied to the SDSS sample.  In the NFGS sample, 121/198 galaxies can be classified using their 
narrow emission-lines according to the \citet{Kewley06} classification scheme.  Of these, 106/121 (88\%) are 
dominated by their star-formation.  The NFGS integrated metallicities have been published by \citet{Kewley04} for several metallicity calibrations.

\section{Stellar Mass Estimates} \label{stellar_masses}

The SDSS stellar masses were derived by \citet{Tremonti04} and \citet{Kauffmann03a} using a combination of $z$-band luminosities and Monte Carlo stellar population synthesis fits to the 4000\AA\ break and the stellar Balmer absorption line H$\delta_{A}$.  The model fits to 
the 4000\AA\ break and H$\delta_{A}$ provide powerful constraints on the star formation history and 
metallicity of each galaxy, thus providing a more reliable indicator of mass than assuming a simple mass-to-light ratio and a \citet{Kroupa01} Initial Mass Function (IMF).  \citet{Drory04} recently compared these spectroscopic masses for $\sim 17000$ SDSS galaxies with (a) masses derived from population synthesis fits to the broadband SDSS and 2MASS colors, and (b) masses calculated from SDSS velocity dispersions and effective radii.  They concluded that the three methods for estimating mass agree to within $\sim 0.2$~dex over the $10^{8} - 10^{12}$~\Msun\ range.     

An alternative method for estimating mass was proposed by \citet{Bell01}.  
Bell et al. used stellar population synthesis models to compute prescriptions for converting optical 
colors and photometry into stellar masses assuming a scaled \citet{Salpeter55} IMF.  This method is useful when near-IR colors are not 
available and spectral S/N is insufficient for reliable 4000\AA\ break and H$\delta_{A}$ measurements.
 We calculate masses for the NFGS galaxies by combining 2MASS J-band magnitudes with the $B-R$ colors (R. Jansen, 2005, private communication).  For all filters, we use 'total' magnitudes, i.e. the integrated light based on extrapolated radial surface brightness fits.  We apply a search radius of 5 arcsec in the 2MASS database, resulting in matches for 85/106 star-forming galaxies.   We calculate stellar masses  using the models of \citet{Bell01}, as parameterised by \citet{Rosenberg05}.   For the comparison between the SDSS and NFGS stellar masses (Section~\ref{aperture}), we assume a Salpeter IMF, 
 and apply factors of 1.82 and 1.43 to the SDSS (Kroupa) and NFGS (scaled Salpeter) stellar 
 masses respectively \citet{Kauffmann03a,Bell01}.

Recently, \citet{Kannappan07} compared stellar masses derived using stellar population synthesis fits to 
the NFGS spectra with masses derived using several methods, including the Bell et al. method.  Kannappan \& Gawiser find that the Bell et al. 2001 stellar mass prescription gives stellar masses that are $\sim 1.5 \times$ the population synthesis approach (see Kannappan \& Gawiser figure 1h).   In the SDSS MZ relation, where stellar 
mass is in log space, a factor of $\sim 1.5$ would result in a shift of $\sim 0.17$~dex.  We consider this shift when comparing the NFGS and SDSS MZ relations (Section~\ref{aperture}).

\section{Metallicity Estimates }\label{Metallicity}

Metallicity calibrations have been developed over $>3$ decades from either theoretical models, empirical calibrations, or a combination of the two.  We apply 10 different metallicity calibrations to the SDSS to investigate the impact of the metallicity calibration on the MZ relation.   We divide the 10 calibrations into four classes; (1) direct, (2) empirical, 
(3) theoretical, and (4) calibrations that are a combination of empirical and theoretical methods.
The empirical, theoretical and combined calibrations all use ratios of strong emission-lines, and are often referred to collectively as "strong-line methods" to distinguish them from the 
``direct" method based on the weak \O4363\  auroral line.

In this paper, we investigate the direct method, five theoretical calibrations, three empirical calibrations , and one "combined" calibration.  We briefly discuss each class of calibration below.  The equations, assumptions, and detailed description of each method that we use are provided in Appendix~\ref{App_calibrations}, and summarized in Table~\ref{calib_table}.  

\subsection{Direct Metallicities}

The most direct method for determining metallicities is to measure the ratio of the \O4363\  auroral line to a lower excitation line such as \OIIIl.  This ratio provides an estimate of the electron temperature of the gas, assuming a classical \HII-region model.  The electron temperature is then converted into a metallicity, after correcting for unseen stages of ionization.   This method is sometimes referred to as the Ionization Correction Factor (ICF), or more commonly, the "direct" method, or the ``\Te" method.  Determining metallicity from the auroral  \O4363\ line is subject to a number of caveats:

\begin{enumerate}
\item The  \O4363\ line is very weak, even in metal-poor environments, and cannot be observed in higher metallicity galaxies without very sensitive, high S/N spectra \citep[e..g.,][]{Garnett04}.  
\item Temperature fluctuations or gradients within high metallicity \HII\ regions may cause electron temperature metallicities to be underestimated by as much as $\sim 0.4$~dex \citep{Stasinska02,Stasinska05,Bresolin06}.  In the presence of temperature fluctuations or gradients, 
\OIII\ is emitted predominantly in high temperature zones where O$^{++}$ is present only in small amounts.  In this scenario, the high electron temperatures estimated from the \O4363\ line are not representative of the true electron temperature in the \HII\ region, leading to systematically low metallicity estimates \citep[see reviews by][]{Stasinska05,Bresolin06}.
\item The \Te\ method may underestimate global spectra of galaxies.  \citet{Kobulnicky99b} found that for low metallicity galaxies, the \Te\ method systematically underestimates the global oxygen abundance of ensembles of \HII\ regions.   
\end{enumerate}

High S/N ratio spectra can overcome the weakness of the \O4363\ line, and alternative auroral lines such as the \NII~$\lambda 5755$,  
\SIII~$\lambda 6312$, and \OII~$\lambda 7325$ lines are observable at higher metallicities than the \O4363\ line  \citep{Kennicutt03,Bresolin04,Garnett04}.  The theoretical investigation by \citep{Stasinska05} predicts that these lines can provide robust metallicities up to $\sim$ solar (\OH$=8.7$; Allende Prieto et al. 2001), but they may underestimate the abundance at metallicities above solar if temperature fluctuations or gradients exist in the nebula.

\subsection{Empirical Metallicity Calibrations}

Because \O4363\ is weak, empirical metallicity calibrations were developed by fitting the relationship
between direct \Te\ metallicities and strong-line ratios for \HII\ regions.  Typical calibrations are based on the optical line ratios  \ratioNIIHa\ \citep{Pettini04},  (\OIIIHb )/(\NIIHa ), \citep[][; hereafter PP04]{Pettini04}, or the ``\R23'' ratio (\ratioR23\ \citep{Pilyugin01,Pilyugin05,Liang07,Yin07}).    PP04 fit the observed relationships between \NIIHa,  \OIIIHb/\NIIHa\  and metallicity for a sample of 137 \HII\ regions.  We refer to the Pettini \& Pagel methods as empirical because 97\% of their sample has \Te\ metallicities. 

\citet[; hereafter P01][]{Pilyugin01} derived an empirical calibration for \R23\ based on \Te-metallicities for a sample of \HII\ regions.  This calibration has been updated by \citet[][; hereafter P05]{Pilyugin05}, using a larger sample of \HII\ regions.  

We refer to strong-line metallicity calibrations that have been calibrated empirically from \Te\ metallicities in \HII\ regions as "empirical methods".   In this paper, we apply the commonly-used empirical calibrations from P01 (revised in P05), and PP04.     These calibrations are described in detail in Appendix~\ref{App_calibrations}, and summarized in Table~\ref{calib_table}.  These empirical calibrations are subject to the same caveats as the \Te-method described above.

\subsection{Theoretical Metallicity calibrations}

The lack of electron temperature measurements at high metallicity led to the development of theoretical 
metallicity calibrations of strong-line ratios using photoionization models.  These theoretical calibrations are commonly and confusingly referred to as "empirical methods".   The use of photoionization models to derive metallicity calibrations is purely theoretical, and the use of the term "empirical" is a misnomer.    We refer to photoionization model-based calibrations as "theoretical methods".   We refer to all calibrations that are based on 
strong line ratios (i.e. including empirical and theoretical methods, but excluding the \Te\ method) as ``strong-line" methods.

Current state-of-the-art photoionization models such as MAPPINGS \citep{Sutherland93,Groves04a,Groves06} and CLOUDY \citep{Ferland98}  calculate the thermal balance at steps through a dusty spherical or plane parallel nebula.   The ionizing radiation field is usually derived from detailed stellar population synthesis models such as Starburst99 \citep{Leitherer99}.   The combination of population synthesis plus photoionization models allows one to predict the theoretical emission-line ratios produced at various input metallicities.   

Photoionization models overcome the temperature gradient problems that may affect \Te\ calibrations at high metallicities because photoionization models include detailed calculations of the temperature structure of the nebula.  However, photoionization models have their own unique set of problems:

\begin{enumerate}
\item  Photoionization models are limited to spherical or plane parallel geometries.
\item  The depletion of metals out of the gas phase and onto dust grains is not well constrained observationally
\item  The density distribution of dust and gas may be clumpy.  This effect is not taken into account with current photoionization models. 
\end{enumerate}

Because of these problems, discrepancies of up to $\sim 0.2$~dex exist among the various strong-line calibrations based on photoionization models \citep[e.g.,][and references therein]{Kewley02a,Kobulnicky04}.  Systematic errors introduced by modelling inaccuracies are usually estimated to be $\sim 0.1 - 0.15$~dex \citep{McGaugh91,Kewley02a}.   These error estimates are calculated by generating large grids of models that cover as many HII region scenarios as possible, 
including varying star formation histories, stellar atmosphere models, electron densities, and geometries.   Differences between the model assumptions and the true HII region ensemble that is observed in a galaxy spectrum are likely to be systematic, affecting all derived metallicities in a similar manner.   

Since systematic errors affect all of the direct, empirical and theoretical methods for deriving metallicities
in high metallicity (\OH $>8.6$) environments, we do not know which method (if any) produces the {\it true} metallicity of an object.  Fortunately, because the errors introduced are likely to be systematic, relative metallicities between galaxies are probably reliable, as long as the same metallicity calibration is used.  We test this hypothesis in Section~\ref{Discussion}.

Many theoretical calibrations have been developed to convert metallicity-sensitive emission-line ratios into metallicity estimates.  Commonly used line ratios include \ratioNIIOII\ \citep[][; hereafter KD02]{Kewley02a} and \ratioR23\ \citep[][; hereafter M91, Z94, and KK04 respectively]{Pagel79,McGaugh91,Zaritsky94,Kobulnicky04}.   In addition to the use of specific line ratios to derive metallicities, theoretical models can be used to simultaneously fit all observed optical emission-lines to derive a metallicity probability distribution, as in \citet[][; hereafter T04]{Tremonti04}.   T04 estimated the metallicity for SDSS star-forming galaxies statistically based on theoretical model fits to the strong emission-lines \OII, \Hb, \OIII, \Ha, \NII, \SII.  

In this paper, we apply the M91, Z94, KK04, KD02, and T04 theoretical calibrations, described in detail in the Appendix~\ref{App_calibrations}.
Many empirical and theoretical metallicity calibrations rely on the double-valued \ratioR23\ line ratio, known as ``\R23".   In Appendix~\ref{R23_break}, we derive the \NIIHa\ and \NIIOII\ values that can be used to break the \R23\ degeneracy in a model-independent way.  

\subsection{Combined Calibration}

Some metallicity calibrations are based on fits to the relationship between strong-line ratios and \HII\ region metallicities, where the \HII\ region metallicities are derived from a combination of 
theoretical, empirical and/or the direct \Te\ method.  For example, the \citet[][; hereafter D02]{Denicolo02} calibration is based on a fit to the relationship between the \Te\ metallicities and the \NIIHa\ line ratio for $\sim 155$ \HII\ regions.  Of these 155 \HII\ regions, $\sim 100$ have metallicities derived using the \Te\ method, and $55$ \HII\ have metallicities estimated using either the theoretical M91 \R23\ method, or an empirical method proposed by \citet{Diaz00} method based on the sulfur lines. 
We refer to calibrations that are based on a combination of methods as ``combined" calibrations.  In this paper, we apply the D02 combined calibration (described in Appendix~\ref{App_calibrations}).

\placetable{calib_table}

\section{The MZ Relation: Metallicity Calibrations}\label{calibrations}

\begin{figure}[!t]
\plotone{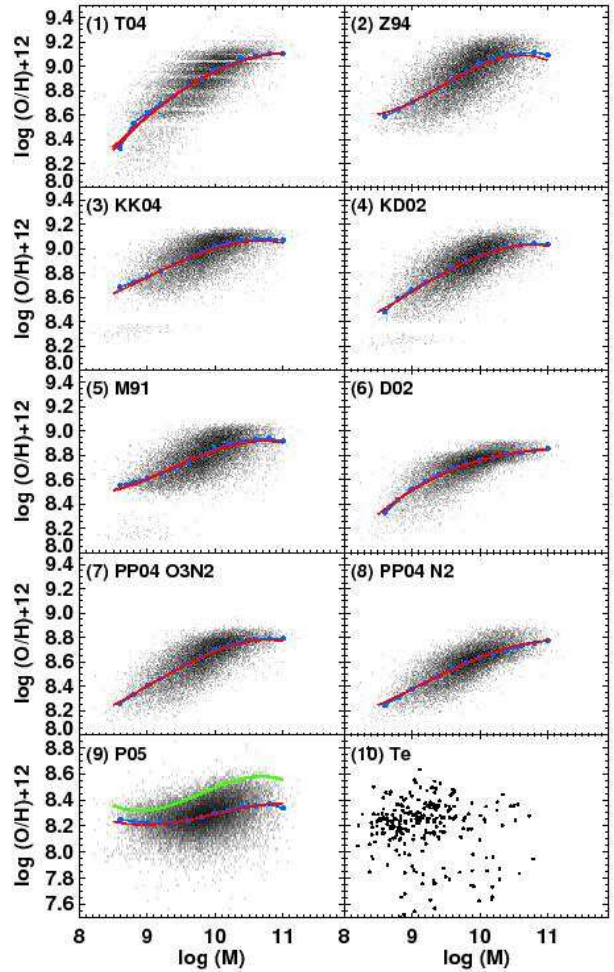} 
\caption{The mass-metallicity relation using the 10 different metallicity calibrations listed in Table~\ref{calib_table}.  The red line shows the robust best-fitting 3rd-order polynomial to the 
data.  The blue circles give the median metallicity within stellar mass bins of $\Delta \log({\rm M}/$\Msun$)=0.2$, 
centered at $\log({\rm M}/$\Msun$)=8.6,8.8,...,11$.   We use the updated calibration of 
P05 given by \citet{Pilyugin05} in panel 9.  The original P01 calibration is shown  as a solid green line in panel 9.   \label{MZ_multiplot}}
\end{figure}

In Figure~\ref{MZ_multiplot} we show the mass-metallicity relation obtained using each of the 10 metallicity calibrations.  There are insufficient galaxies in the SDSS with \OIII~$\lambda 4363$ detections to determine an MZ relation using the \Te\ metallicities.  For the strong-line methods (i.e. all 
methods except the direct \Te\ method),
the red line shows the robust best-fitting 3rd-order polynomial to the 
data.  The blue circles give the median metallicity within masses of $\log({\rm M})=0.2$~\Msun, 
centered at $\log({\rm M})=8.6,8.8,...,11$~\Msun.    Both methods of characterizing the shape of the 
MZ relations produce similar results.   The parameters of the best-fit polynomials and the rms residuals of the fit are given in Table~\ref{MZ_table}.

The different strong-line calibrations produce MZ relations with different shapes, y-axis offsets, and scatter.   T04 interpret the flattening in the MZ relation above 
stellar masses $\log({\rm M})>10.5$~\Msun\ in terms of efficient galactic scale winds that remove metals from the galaxies with masses below $\log({\rm M})< {10.5}$~\Msun.  A similar flattening is observed for the majority of 
the theoretical techniques.  However, the MZ relations calculated using metallicity calibrations based on \NIIHa\ (D02 and PP04 N2) flatten at lower stellar masses $\log({\rm M})\sim10$ because the \NIIHa\ line ratio becomes insensitive to metallicities for log(\NIIHa)$\gtrsim -1$ 
(or \OH$\gtrsim 8.8$ in the D02 or PP04 \NIIHa-based metallicity scale).  The \NIIHa\ calibrations 
cannot give metallicity estimates above \OH$\gtrsim 8.8$, even if the true metallicity is higher than \OH$>8.8$.

The P05 empirical method  \citep{Pilyugin05} is relatively flat  for all stellar masses; between $8.5 \leq \log({\rm M}/{\rm M}\odot)\leq11$, the metallicity rises only $\sim 0.2$~dex on average.   The majority of the \HII\ regions used by P05 have \Te\ metallicities that are based on the \OIII~$\lambda4363$ line.  Because the \OIII~$\lambda4363$ line may be insensitive to (or saturate at) a metallicity \OH$\sim 8.6$, 
the P05 calibration may give a weak MZ relation for the SDSS.  Interestingly, the original P01 calibration (green line in panel (9) of Figure~\ref{MZ_multiplot}) gives a steeper MZ relation than the updated calibration (P05; red and blue lines).  The updated P05 relation also produces lower absolute metallicities by $\sim 0.2$~dex compared with the original P01 method, as pointed out by \citet{Yin07} 
in their comparison between P01, P05, and T04 metallicities.
 This change may be caused by the different \HII-region abundance sets that were used to calibrate the original P01 method and the updated version in P05.

The direct \Te\ method is available for only 546/27,730 (2\%) of the galaxies in our SDSS sample.   The \OIII~$\lambda 4363$ line is weak and is usually only observed in metal-poor galaxies.  The SDSS catalog contains very few metal-poor galaxies because they are intrinsically rare, compact and faint \citep[e.g.,][]{Terlevich91,Masegosa94,VanZee00}.   Panel 10 of Figure~(\ref{MZ_multiplot} shows that 
a total of 477 \Te\ metallicities is insufficient to obtain a clear MZ relation.  Because we are unable to fit an MZ relation using \Te\ metallicities, we do not consider the \Te\ method further in this work. 

The scatter in the MZ relation is large for all metallicity calibrations; the rms residual about the line of 
best-fit is 0.08 - 0.13.  The cause of the scatter in the MZ relation is unknown.   Our comparison 
between the different metallicity calibrations shows that differing ionization parameter among 
galaxies does not cause or contribute to the scatter.  The ionization parameter is explicitly calculated and taken into account in some metallicity diagnostics (KD02, KK04, M91), but we do not see a reduction in scatter for these methods.  A full investigation into the scatter in the MZ relation will be 
presented in Ellison et al. (in prep).

\begin{figure}[!t]
\plotone{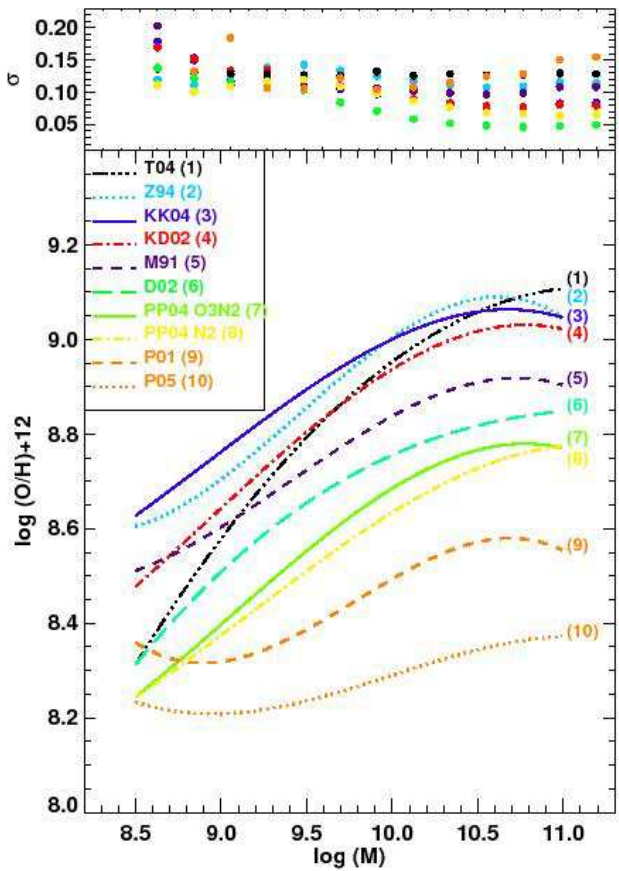} 
\caption{The robust best-fit mass-metallicity relations calculated using the  different metallicity calibrations listed in Table~\ref{calib_table}, except the \Te-method.  The top panel shows the rms scatter in metallicity about the best-fit relation for each calibration in 0.1~dex bins of stellar mass. The y-axis offset, shape, and scatter of the MZ relation differs substantially, depending on which metallicity calibration is used.  
\label{MZ_median}}
\end{figure}

\placefigure{MZ_multiplot}
\placefigure{MZ_median}
\epsscale{1.2}

We directly compare the best-fit MZ curves for the 9 strong-line calibrations in Figure~\ref{MZ_median}, including both P01 and P05.
The top panel shows the rms scatter in metallicity about the 
mean in mass bins of width $\Delta \log({\rm M}/$\Msun$)=0.2$.  The major difference between the MZ curves is their position along the y-axis.    The curves with the largest y-intercept are all photoionization model based (KK04, Z94, KD02, T04, M91).  Among these photoionization model metallicities, the agreement is $\sim 0.2$~dex.  This agreement is within 
the margin of error typically cited for these calibrations ($\sim 0.1 - 0.15$~dex for each calibration).   Some calibrations consistently agree to within 0.1~dex (e.g., KK04 and Z94; KD02 and M91).  Comparisons between metallicities calculated using these consistent methods, such as KD02 and M91, are likely to be reliable to within 0.1~dex.  However, comparisons between methods that show large disagreement (such as KK04 and P05) will be contaminated by the large systematic discrepancy between the calibrations.

 The lowest curves in Figure~\ref{MZ_median} are the MZ relations derived using the empirical methods (i.e. P01, P05, and the two PP04 methods).  These empirical methods are calibrated predominantly via fits of the relationship between strong-line ratios and \HII\ region \Te\ metallicities.   There is considerable variation among the y-intercept of these \Te-based MZ relations; the P05 method gives metallicities that are $\sim 0.4$~dex below the PP04 methods at the highest masses, despite the fact that both methods are predominantly based on \HII\ regions with \Te-metallicities.  At the lowest stellar masses, this difference disappears.  The difference between the empirical methods may be attributed to the different \HII-region samples used to derive the calibrations.  At the highest metallicities, the PP04 methods utilize four \HII-regions with detailed theoretical metallicities.  These detailed theoretical metallicities may overcome the saturation at \OH$\sim 8.6$ suffered by \OIII~$\lambda 4363$ \Te\ metallicities.   The P05 calibration includes some \HII\ regions with metallicities estimated with the alternative auroral \NII~$\lambda 5755$ line from \citet{Kennicutt03}.  The inclusion of these \NII~$\lambda 5755$ metallicities may overcome the \OIII~$\lambda 4363$ saturation problem.  However, \citet{Stasinska05} suggest that the use of the \NII~$\lambda 5755$ line in dusty nebulae will still cause \Te\ metallicities to be underestimated when the true metallicity is above solar.   Our SDSS sample has a mean extinction of E(B-V)$\sim0.3$, or $A_{V}\sim1$.   The extinction is a strong function of stellar mass; for the largest stellar masses (M$>10^{10.5}$\Msun), the mean extinction is large E(B-V)$\sim0.5$, or $A_{V}\sim1.6$.  Clearly, dust is important in SDSS galaxies, particularly at the highest stellar masses where the largest discrepancies exist between the theoretical methods and the P05 \Te-methods.   

In addition to the large difference in y-intercept between the different metallicity calibrations, Figure~\ref{MZ_median} shows that the slope and turn-over of the MZ relation
depend on which calibration is used.   Therefore, it is essential to compare MZ relations that have been 
calculated using the same metallicity calibration.   In the following section, we derive conversions that 
can be used to convert metallicities from one calibration into another.

\section{Metallicity Calibration Conversions}\label{Conversions}

Comparisons between MZ relations for galaxies in different redshift ranges are non-trivial.
Different suites of emission-lines are available at different redshifts, necessitating the use of different metallicity calibrations.  Because of the strong discrepancy in absolute metallicities between different calibrations, the application of different calibrations for galaxies at different redshifts may  mimic or hide evolution in the MZ relation with redshift, depending on which calibrations are used.  Because the metallicity discrepancies are systematic, we can fit the relationship between the 
different metallicity calibrations in order to remove the systematic discrepancies and obtain comparable metallicity measurements for different redshift intervals.  

\begin{figure}[!t]
\plotone{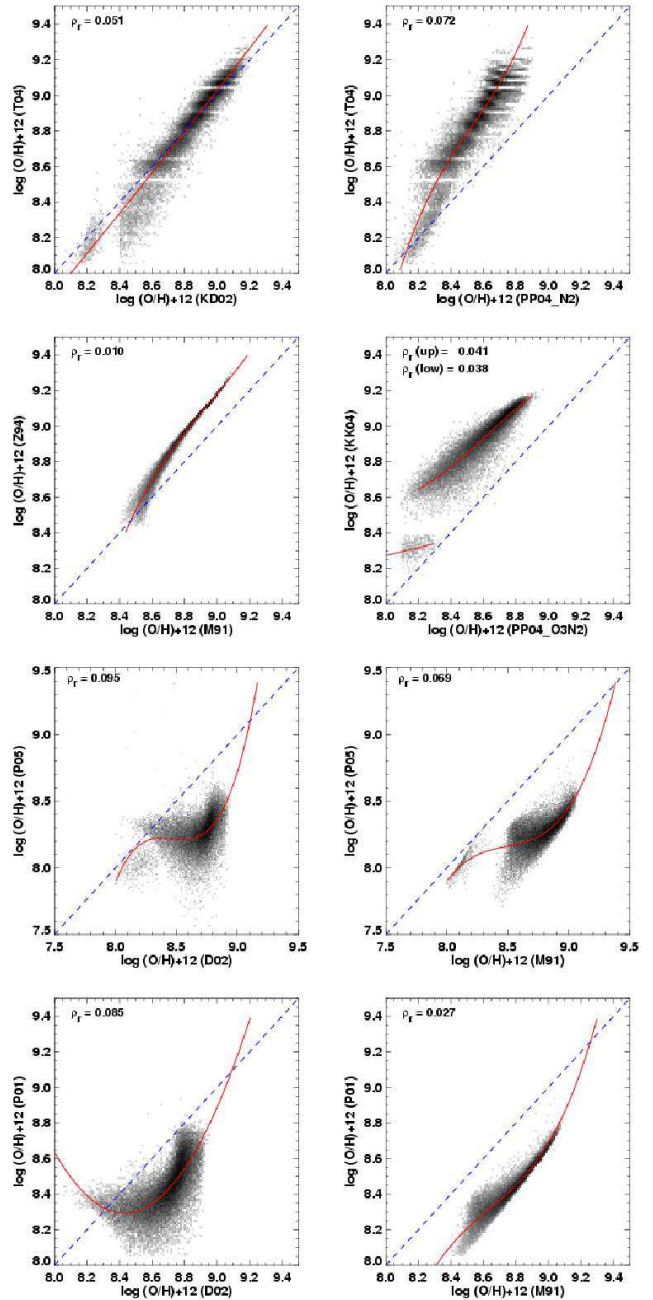} 
\caption{Examples of the relationship between different metallicity calibrations.  The robust 2nd order polynomial of best fit is shown as a red solid line.  The 1:1 line (blue dashes) shows where the metallicities would lie if the calibrations agree.  The robust equivalent to the standard deviation of the fit ($\rho_r$) are shown for each plot.  This figure illustrates the typical variation in scatter and shape between different metallicity calibrations.  Figures showing the relations between all 9 strong-line metallicity calibrations are available at http://www.ifa.hawaii.edu/$\sim$kewley/Metallicity.
\label{calib_comp1}}
\end{figure}

We calculate conversion relations between the strong-line metallicity calibrations by plotting each calibration against the remaining 8 calibrations and fitting the resulting metallicity-metallicity distribution with a robust polynomial fit.   We refer to these metallicity-metallicity plots as Z-Z plots.   Rows 1-3 of Figure~\ref{calib_comp1} give six representative examples of SDSS Z-Z plots for the strong-line calibrations.    The Z-Z plots between all nine strong-line calibrations for various S/N cuts are available at {\it http://www.ifa.hawaii.edu/$\sim$kewley/Metallicity}.  The blue dashed 1:1 line shows where the Z-Z distribution would lie if the two calibrations agree.     The robust best-fit polynomial is shown in red, and $\rho_r$ gives the robust equivalent to the standard deviation of the fit. Small values of  $\rho_r$ indicate a reliable fit to the data.   

The majority of the Z-Z relations are close to linear and are easily fit by a 1st, 2nd or 3rd order robust polynomial.  However, the P05 calibration produces a very non-linear relation with a large scatter when plotted against all other metallicity calibrations.  These non-linear relations are not easily fit even with a robust 3rd order polynomial and we cannot provide conversions that will reliably convert to/from the P05 method.   For comparison, the bottom row of Figure~\ref{calib_comp1} shows the same plots calculated with the original P01 calibration.  Although the scatter is less severe in these plots, the relations between P01 and other diagnostics remain non-linear and are not easily fit with a robust 3rd order polynomial.

\placefigure{calib_comp1}

For all other diagnostics, the metallicities or metallicity relations can be converted into any other calibration scheme, using

\begin{equation}
y = a + bx +cx^2 + dx^3
\end{equation}

\noindent
where $y$ is the "base" or final metallicity in \OH\ units, $a-d$ are the 3rd order robust fit coefficients given in Table~\ref{conv_table}, and $x$ is the original metallicity to be converted (in \OH\ units).  For Z-Z relations where a 2nd order polynomial produces a lower $\rho_r$ than a 3rd order polynomial fit, $d$ is zero.

 The conversion coefficients given in Table~\ref{conv_table} are based on the fit order that produces the lowest $\rho_r$ value in our sample.  Some \R23\ calibrations require two fits; one 2nd or 3rd order fit for the upper \R23\ branch and one linear fit for the lower branch.  In these cases, the coefficents of the upper and lower branch fits are listed in Table~\ref{conv_table} as left and right columns, respectively.

In Table~\ref{conv_table}, we give the range in $x$ over which our calibrations are valid.  
 Our polynomial fits are only tested within these ranges and may not be suitable for converting lower or higher metallicities into another scheme outside these limits.  We provide worked examples for the use of our conversions in Appendix~\ref{worked_examples}.   
 
Figure~\ref{MZ_median_fix} shows the application of our strong-line conversions to the best-fit MZ relations in Figure~\ref{MZ_median}, excluding P05.   The calibration shown for each panel represents the ``base" (final) calibration into which all other MZ curves have been converted.   The remaining discrepancy between the converted MZ relation and the base MZ relation is an indicator of both the scatter in our Z-Z plots and how well the Z-Z relations are fit by a robust polynomial.    In Table~\ref{dev_table}, we give the mean residual discrepancy between the converted MZ relations and the base MZ relation.   

Our conversions reach agreement between the MZ relations to within $\sim 0.03$~dex on average. 
The most reliable base calibrations are those with the smallest residual discrepancies.  The residual discrepancies differ because some Z-Z relations have less scatter and/or are more easily fit by a simple polynomial.   The KK04, M91, PP04 O3N2, and KD02 methods have the smallest residual discrepancies and are therefore the most reliable base calibrations to convert other metallicities into.

\epsscale{1.2}

\epsscale{1.2}
\begin{figure}[!t]
\plotone{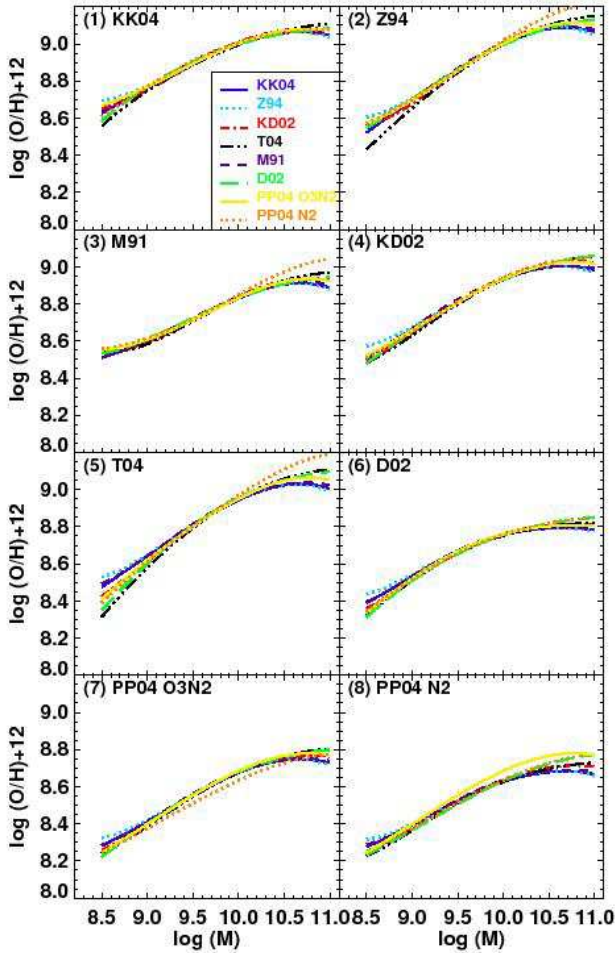}  
\caption{The robust best-fit mass-metallicity relations calculated using the 8 different metallicity calibrations listed in Table~\ref{calib_table}, converted into the base metallicities shown using our conversions from Table~\ref{conv_table}.  
\label{MZ_median_fix}}
\end{figure}

\section{The MZ Relation: AGN removal methods}\label{AGN}

The nebular emission line spectrum is sensitive to the
hardness of the ionizing EUV radiation.     Metallicities calculated from spectra that contain a 
significant contribution from an AGN may be spurious because the commonly-used 
metallicity calibrations are based on the assumption of a stellar ionizing radiation field.
The standard optical diagnostic diagrams for classification were first
proposed by \citet{Baldwin81}, based on the line ratios \NIIHa\ vs \OIIIHb, \SIIHa\ vs \OIIIHb, and \OIHa\ vs \OIIIHb.  This classification scheme was revised by \citet{Osterbrock85} and %
\citet[][; hereafter VO87]{Veilleux87} who used a combination of AGN and starburst samples with photoionization 
models to derive a classification line on the diagnostic diagrams to separate AGN from starburst 
galaxies.  Subsequently, \citet[][; hereafter Ke01]{Kewley01a} developed a purely theoretical ``maximum starburst line" 
line for AGN classification using the standard diagrams. This theoretical scheme provides an improvement on the 
previous semi-empirical classification by producing a more consistent classification line from diagram to diagram that 
significantly reduces the number of ambiguously classified galaxies.     The ``maximum starburst line" defines the
maximum theoretical position on the diagnostic diagrams that can be attained by pure star formation
models.  According to the Ke01 models, galaxies lying above the maximum starburst line are dominated by AGN activity and objects lying below the line are dominated by star formation.   

Although objects lying below the maximum starburst line are likely to be dominated by star formation, they may contain a small contribution from an AGN.   We calculate the maximum AGN contribution that would allow a galaxy to be classified as star-forming with the Ke01 line on all three standard diagnostic diagrams using theoretical galaxy spectra.  Our AGN model is based on the \OH$=8.9$ dusty radiation-pressure dominated models by \citep{Groves04a}.  We use a typical AGN ionization parameter of $\log(U)=-2$ and a power-law index of $\alpha=-1.4$.   We investigate the suite of starburst models from 
\citet{Kewley02a} and \citet{Dopita00}.  The starburst model that allows the maximum contribution from an AGN while remaining classified as star-forming is zero-age instantaneous burst model with ionization parameter $q=1\times10^7$~cm/s and metallicity \OH$=8.9$ by \citep{Kewley02a,Dopita00}.   The AGN contribution in this model is  $\sim 15$\%.

We use this model to calculate the effect of a 15\% AGN contribution to the metallicity-sensitive emission-line ratios. 
The AGN model contributes substantially to the \OIIIHb\ line ratio but has only a minor effect on the \NIIOII\ ratio.  Therefore, the effect of 
an AGN contribution of 15\% is small ($\leq 0.04$~dex) on metallicities calculated using the \NIIOII\ ratio \citep{Kewley02a}, but larger ($0.1-0.2$~dex) on metallicities calculated with calibrations containing \OIII\ \citep[e.g.,][]{McGaugh91,Zaritsky94,Kobulnicky04}.   

Recently, \citet[][; hereafter Ke06]{Kewley06} defined a new classification scheme based on all three diagnostic diagrams that separates pure HII region-like galaxies from HII-AGN composites, Seyferts, and galaxies dominated by low ionization emission line regions (LINERs).  This new classification scheme includes an empirical shift applied by \citet[][; hereafter Ka03]{Kauffmann03a} to the Ke01 line for the \NIIHa\ vs \OIIIHb\ diagnostic.  This shift provides a more stringent removal of objects containing AGN, and we recommend its use for metallicities calculated using \R23.  

\begin{figure}[!h]
\plotone{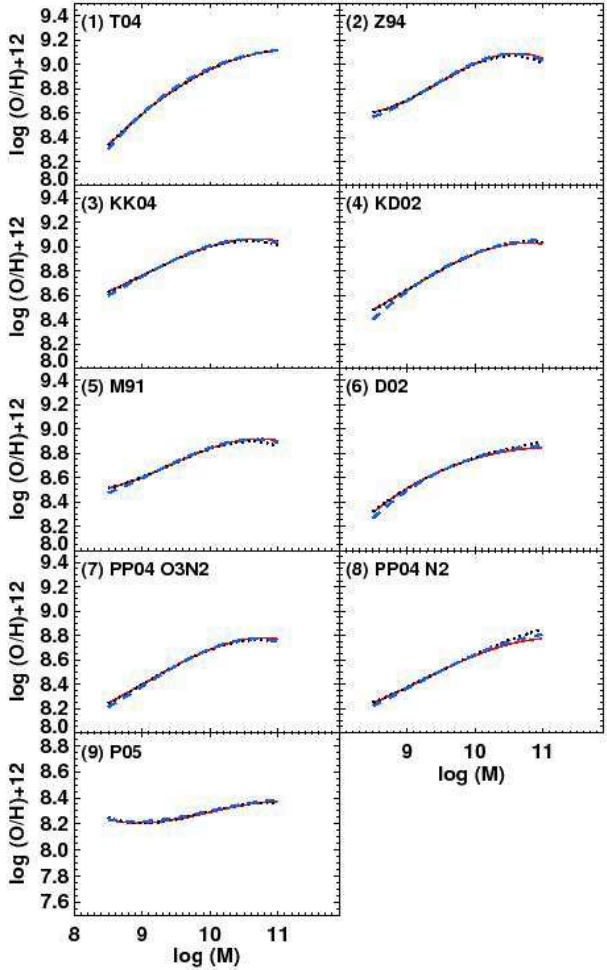} 
\caption{Comparison between the mass-metallicity relations for three different methods of 
AGN removal: \citet{Kewley06} (red solid line), \citet{Kewley01a} (black dotted line), and \citet{Veilleux87} (VO87; blue dashed line).   The method of AGN removal has little effect on the MZ relation, except at low stellar masses here the VO87 method gives a flatter slope.
\label{MZ_class}} 
\end{figure}

We investigate whether the AGN classification scheme affects the shape of the MZ relation in Figure~\ref{MZ_class}.  For each metallicity calibration, we show the MZ relation for the three classification schemes Ke01 (black dotted line), Ke06 (red solid line) and VO87 (blue dashed line).  
These three classification schemes define 89\%, 84\%, and 76\% of our SDSS sample as star-forming, 
respectively.   There is negligible difference ($<0.05$~dex) among the SDSS MZ relations for the three classification schemes.   We note that the contribution from an AGN may be more important for samples that contain a larger fraction of HII-AGN composite galaxies, or galaxies at high redshift where limited sets of emission-lines limit the methods for AGN removal.  For these cases, we recommend the use of either the 
KD02 \NIIOII\ metallicity calibration (useful for log(\NIIOII)$> -1.2$), the PP04 \NIIHa\ calibration, or the 
D02 \NIIHa\ calibration.  None of these three calibrations depend on the AGN-sensitive \OIIIHb\ line ratio.

\placefigure{MZ_class}

\section{The MZ Relation: Aperture Effects} \label{aperture}

\begin{figure}[!ht]
\plotone{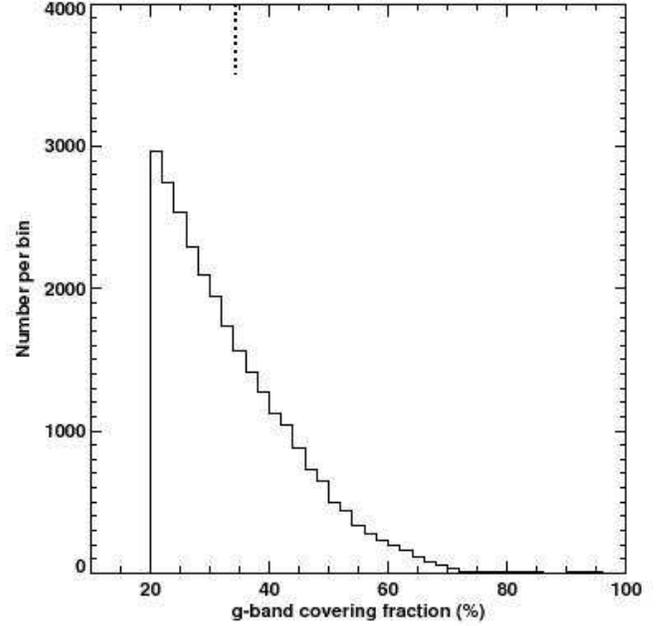} 
\caption{The distribution of g'-band fiber covering fractions in our SDSS sample.  Our sample was chosen to have covering fractions $>20$\%.  The dotted line at the top of the figure indicate the 
median (34.2) covering fraction of our sample.
\label{cov_frac}}
\end{figure}

Our SDSS sample was selected with g'-band covering fractions $>20$\% because this value is the minimum covering fraction required for metallicities to approximate the global values \citep{Kewley05a}.   A covering fraction of $20$\% corresponds to 
a median redshift of  $z\sim 0.04$ which is the lower redshift limit used by T04 for their SDSS MZ relation work.   The median g'-band aperture covering fraction of our sample is only $\sim$34\%, although the range of g'-band covering fractions is 20-80\% (Figure~\ref{cov_frac}).    

\placefigure{cov_frac}

Strong metallicity gradients exist in most massive late-type spirals; \HII\ region metallicities decrease by an order of magnitude from the inner to the outer disk \citep[see e.g.,][for a review]{Shields90}.  These gradients may cause substantial differences between the nuclear and global metallicities.  \citet{Kewley05a} investigate the effects of a fixed size aperture on spectral properties for a large range of galaxy types and luminosity.   They conclude that minimum covering fractions larger than $20$\% may be needed at high luminosities to avoid aperture effects.  Therefore the SDSS MZ relation may be affected by the fixed size aperture at the highest luminosities or stellar masses.   

\begin{figure}[!h]
\plotone{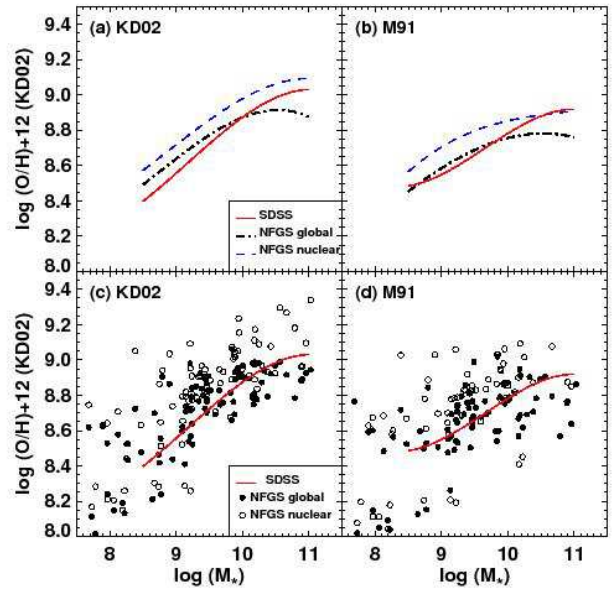} 
\caption{(Top Panel) Comparison between the robust best-fit SDSS Mass-Metallicity relation (red solid line) and  the best-fit relations to the Nearby Field Galaxy Survey (NFGS) nuclear (blue dashed line) and global (black dot-dashed line) relations.  Metallicities were calculated using both KD02 (left panel) and M91 (right panel) calibrations, and stellar masses are given assuming a Salpeter IMF.  (Bottom Panel) Comparison between the  SDSS Mass-Metallicity relation (red solid line) and the NFGS global (filled circle) and NFGS nuclear (unfilled circle) data.  At high stellar masses 
(${\rm M}_{*} > 10^{10}$\Msun), the SDSS metallicities are $\sim 0.1 - 0.15$~dex larger than NFGS global metallicities at the same stellar mass.  
\label{MZ_ap}}
\end{figure}

To investigate the effect of the small median SDSS covering fraction on the MZ relation, in Figure~\ref{MZ_ap}, we compare the SDSS MZ relation (red solid line) with the nuclear (black dot-dashed line) and global (blue dashed line) MZ relations of the Nearby Field Galaxy Survey (NFGS).     We show the KD02 metallicity calibration (left panel) and the M91 calibration (right panel) for all datasets.   Similar results are obtained with the other strong-line methods.
The SDSS MZ relation lies in-between the NFGS nuclear and global relations at high stellar masses (${\rm M}> 10^{10}$\Msun).  The NFGS global MZ relation flattens at a metallicity that is $\sim0.1 - 0.15$~dex smaller than the metallicity at which the SDSS relation flattens.   This difference is not 
caused by metallicity calibration errors because the difference in upper turn-off is observed with all strong-line metallicity calibrations.   Furthermore, the difference of $\log({\rm M})= 0.17$~dex  between the \citet{Bell01} stellar mass relation and the SDSS Bruzual \& Charlot model stellar masses cannot account for the difference between the SDSS and global NFGS MZ relations.

\placefigure{MZ_ap}

The difference between the SDSS and NFGS {\it nuclear} and MZ relations is probably driven by two 
factors: (1) fixed-size aperture differences and (2) different surface brightness profiles. The NFGS nuclear sample has a smaller mean covering fraction than the SDSS sample ($\sim10$\% c.f. $\sim34$\%), giving higher nuclear metallicities in the NFGS than for SDSS galaxies with the same stellar mass.  In addition, the NFGS and SDSS samples have different surface brightness profiles (traced by their half-light radii).   The NFGS sample has a slightly smaller mean half light radius than our
SDSS sample ($\sim 3.0$~kpc c.f. $\sim 3.4$~kpc respectively).  Ellison et al. (in prep.) show that for the SDSS,  galaxies with small g'-band half light radii (i.e more concentrated emission) have higher metallicities at a given mass than galaxies with large half light radii (more diffuse emission).  

The difference in half-light radii between the SDSS and NFGS samples can not explain the difference between SDSS galaxies and {\it global} NFGS MZ relations at high stellar masses ${\rm M}_{*} > 10^{10}$\Msun because (a) the larger mean half-light radii of the SDSS sample would bias the SDSS towards {\bf lower} metallicities than the NFGS (Ellison et al. in prep), and (b) the NFGS global aperture covering fraction ($\sim 82$\%) captures most of the NFGS B-band emission.  The half-light radius is irrelevant when the spectrum captures 100\% of the B-band light.

We conclude that a g'-band covering fraction of $\sim 20$\% (or lower redshift limit of $z=0.04$) is insufficient for avoiding aperture bias in SDSS galaxies with stellar masses ${\rm M}_{*} > 10^{10}$\Msun.     The mean covering fraction for ${\rm M}_{*} > 10^{10}$\Msun\ galaxies is $30.6\pm0.1$\%.   A larger mean covering fraction is required to obtain a reliable MZ relation at ${\rm M}_{*} > 10^{10}$\Msun.

\section{Discussion }\label{Discussion}

We have investigated the effect of metallicity calibrations, AGN removal schemes, and a fixed-size aperture on the MZ relation.  The choice of metallicity calibration has the strongest effect on the MZ relation.   The choice of metallicity calibration changes the y-intercept of the MZ relation significantly; the discrepancy between the metallicity calibrations is as large as $0.7$~dex at the highest stellar masses.  This discrepancy corresponds to a difference of 0.5 to $2.6 \times$~solar at the peak metallicity of our SDSS sample.

The existence of a $\sim 0.4-0.5$~dex discrepancy between the T$_e$ and theoretical metallicities is well known \citep{Stasinska02,Kennicutt03,Garnett04}.   Our results show that the discrepancy is larger than previously thought. This discrepancy is often interpreted as an unknown problem with the photoionization codes used to calibrate the strong line ratios with metallicity \citep[][]{Kennicutt03,Garnett04,Tremonti04}.   However, recent investigations indicate that the \Te\ methods may underestimate the metallicity when temperature fluctuations or gradients exist within the emission-line nebulae \citep{Stasinska05,Bresolin06}.  These fluctuations, and hence the effect on \Te\  are expected be the strongest at the highest metallicities.   We conclude that for the metallicities spanned by the SDSS sample, it is not possible to know which (if any) metallicity calibration is correct.  
Until the metallicity discrepancies are resolved, only {\it relative} metallicity comparisons should be made.   

Relative metallicity comparisons rely on the ability of strong-line calibrations to consistently reproduce 
the metallicity difference between any two galaxies.   For example, if two galaxies have metallicities of 
\OH$=8.4$ and \OH$=8.9$ using one metallicity calibration, the difference in relative metallicities (0.5~dex) should be the same using any other metallicity calibration, even if the absolute metallicities differ from one calibration to another.  We test how well the strong-line metallicity calibrations maintain relative metallicities by selecting $30,000$ random sets of two galaxies from our SDSS sample.  We measure the relative metallicity difference between the two galaxies from each set for each metallicity calibration.  We give the mean difference in relative metallicity and rms residuals in Table~\ref{relative_rms_table}.  The mean difference in relative metallicity is $<0.07$~dex for all strong-line metallicity calibrations.  The rms scatter is $\lesssim 0.15$~dex for all calibrations.  The P05 method gives more discrepant relative metallicities to the other strong-line methods, with relative metallicity differences between $0.08 - 0.14$~dex rms (c.f. 0.02 - 0.11~dex rms).  The best agreement between relative metallicities occurs between the three theoretical \R23\ calibrations (M91, Z94, KK04), with relative metallicities agreeing to within $0.02 - 0.05$~dex rms.   The small difference and rms residuals in relative metallicities for all 9 strong-line calibrations indicates that comparisons between galaxy or \HII-region metallicities can be reliably made to within $\sim 0.15$ dex, as long as the same base metallicity calibration is used for galaxies or \HII\ regions.    Our metallicity conversions aid relative metallicity comparisons between different samples of galaxies at different redshifts by empirically removing the discrepancy between each metallicity calibration.  In practice, if relative metallicity differences between galaxies or between samples is $\lesssim 0.15$~dex, we recommend the use of two or more metallicity calibrations to verify that any difference observed is real, and not introduced by the metallicity calibration applied.

The SDSS sample is insufficient for determining the cause(s) of the metallicity discrepancy problem.
Several ongoing investigations into the metallicity discrepancies may help solve this problem in the near future.  These investigations include tailored photoionization models, high S/N spectroscopy of luminous stars in the Milky Way and nearby galaxies, metal recombination lines, IR fine structure lines, and temperature fluctuation studies.

\citet{Garnett04b} applied tailored photoionization models to optical and infrared spectra of the \HII\ region CCM 10 in M51.  They
found that the CCM 10 metallicity derived from the electron temperature using the infrared \OIII~88$\mu$m line agrees with the theoretical metallicity computed with the latest version of the CLOUDY v90.4 photoionization code \citep{Ferland98}.  This theoretical metallicity is a factor of 2 smaller than the metallicity calculated with the previous version of CLOUDY (v. 74) that uses older atomic data.   Clearly the optical emission-line strengths in the photoionization models are very sensitive to the atomic data used.  However, this sensitivity cannot explain the discrepancy observed in Figure~\ref{MZ_median} because T04 used the same version of Cloudy as Garnett et al.  In spite of the use of modern photoionization models with more accurate atomic data, the T04 MZ relation lies significantly higher than the methods utilizing \Te\ metallicities (P05,D02,PP04).   \citet{Mathis05} used Monte Carlo photoionization models to show that different density distributions are not a significant source of error on the theoretical abundances.   Recently, \citet{Ercolano07} used new 3D photoionization codes to investigate the effect of multiply non-centrally located stars on the temperature and ionization structure of \HII\ regions.  Ercolano et al. suggest that the geometrical distribution of ionizing sources may affect the metallicities derived using theoretical methods.   Further theoretical investigations into the model assumptions, as well as tailored photoionization model fits to multi-wavelength data of spatially resolved star-forming regions may yield clues into whether the theoretical models contribute to the metallicity discrepancy.
 
High S/N spectroscopy with 8-10m telescopes can now provide metallicities for luminous stars and planetary nebulae in nearby galaxies that can be compared with \HII\ region metallicities \citep[see][for a review]{Pryzbilla07}.  \citet{Urbaneja05} analysed the chemical composition of B-type supergiant stars in M33.  They find that the supergiant metallicities agree with \HII\ region abundances derived using the \Te- method.  Similar results were obtained for local dwarf galaxies \citep{Bresolin06c}, however other investigations require a correction for electron temperature fluctuations before agreement can be reached \citep{Simon06}.
  
Metal recombination lines provide a promising independent baseline for metallicity measurements because metal recombination lines depend only weakly on \Te \citep[see e.g.,]{Bresolin06}.  Metal recombination lines are weak, but they have been observed in \HII\ regions in the Milky Way \citep{Esteban04,Esteban05,Garcia05,Garcia06}, and recently in nearby galaxies \citep{Esteban02,Peimbert05b}.   Recombination methods typically agree with theoretical methods 
 \citep[e.g.,][]{Bresolin06}, and predict larger metallicities (by 0.2-0.3~dex) than the \Te\ method.  When the \Te\ metallicities are corrected for electron temperature fluctuations, agreement is reached between recombination and \Te\ methods 
 \citep{Peimbert05b,Garcia05,Garcia06,Peimbert06,Lopez07}.   
 
Measurements of  electron temperature fluctuations in nearby \HII\ regions can resolve the  disagreement between strong-line theoretical methods, and electron temperature methods \citep{Garcia06,Bresolin07} in most cases \citep[see however][]{Hagele06}. More electron temperature measurements are needed to verify these results, particularly for high metallicity \HII\ regions where electron temperature fluctuation measurements are lacking.
 
 Despite these promising investigations, the metallicity discrepancy problem remains unsolved at the present time.
 Until the metallicity discrepancy problem is resolved, absolute metallicity values should be used with caution.   In Table~\ref{relative_rms_table}, we show that the metallicity calibrations maintain relative metallicities better than $\sim 0.15$~dex rms, with the majority of theoretical methods maintaining relative metallicities within $0.04-0.1$~dex rms.   Therefore, studies of relative metallicity differences, such as comparisons between different galaxy samples, or between individual galaxies or \HII\ regions, can be reliably made.  If the size of the differences observed between different samples or different objects is $\sim 0.15$~dex or less, we recommend the use of at least two independent calibrations to verify that the difference is calibration-independent.  The KD02 and PP04 methods give (a) low residual discrepancies in relative metallicities,  and (b) low residual discrepancy after other metallicities have been converted into these two methods.  For the metallicity range of the SDSS sample, the KD02 and PP04 N2 calibrations are also independent of small contributions from an AGN.  Because the KD02 and PP04 O3N2 methods maintain robust relative metallicities and are good base calibrations, we recommend the use of these two methods when deriving relative metallicities.   
 
\section{Conclusions} \label{Conclusions}

We present a detailed investigation into the mass-metallicity relation for 27,730 star-forming galaxies from the 
Sloan Digital Sky Survey.  We apply 10 different metallicity calibrations to our SDSS sample, including theoretical photoionization calibrations and empirical \Te\ method calibrations.   We investigate the effect of these  metallicity calibrations on the shape and y-intercept of the mass-metallicity relation.   Using 30,000 galaxy sets sampled randomly from our SDSS sample, we investigate how well the 9 strong-line calibrations maintain relative metallicities.   We find that:

\begin{itemize}
\item  The choice of metallicity calibrations has the strongest effect on the MZ relation.  The choice of metallicity calibration can change the y-intercept of the MZ relation by up to 0.7~dex.   Until this metallicity discrepancy problem is resolved, absolute metallicities should be used with extreme caution.
\item There is considerable variation in shape and y-intercept of the MZ relations derived using the empirical methods.  We attribute this variation to the different \HII\ region samples used to derive the empirical calibrations.
\item The relative difference in metallicities is maintained to an accuracy of \OH$\sim 0.02 - 0.1$~dex for 9/10 calibrations, and to within \OH$\sim 0.15$~dex for all 9 strong-line calibrations.   For relative metallicity studies where the difference between targets or between samples is $\lesssim 0.15$~dex, we recommend the use of at least two different calibrations to check that any result is not  
caused by the metallicity calibrations.   
\end{itemize}

We use robust fits to the observed relationship between each metallicity calibration to derive new conversion relations for converting metallicities calculated using one calibration into metallicities of another, "base" calibration.   We show that these conversion relations successfully remove the strong discrepancies observed in the MZ relation between the different calibrations.   Agreement is reached to within 0.03~dex on average.

We investigate the effect of AGN classification scheme and fixed-size aperture on the MZ relation. 

\begin{itemize} 
\item AGN classification methods have a negligible effect on metallicities derived using \NIIOII\ or \NIIHa.  AGN classification can affect metallicities derived with \R23\ by $\lesssim 0.15$~dex.   For the SDSS sample, AGN classification methods make negligible difference in the shape or y-intercept of the MZ relation.  For samples containing a larger fraction of starburst-AGN composite galaxies, or samples where AGN removal is not possible, we recommend the use of \NIIOII\ or \NIIHa\ metallicity diagnostics.
\item The median g'-band aperture covering fraction of our SDSS sample is 34.2\%.  This covering fraction is insufficient for metallicities to represent global values at high masses (M$>10^{10}$\Msun).   The Nearby Field Galaxy global MZ  is $0.1 - 0.15$~dex lower than the SDSS MZ relation at M$>10^{10}$\Msun.  Therefore, the metallicity at which the SDSS MZ relation turns over is dependent on both the choice of metallicity calibration, and on the aperture size.
\end{itemize}

We recommend that metallicities be converted into either the \citet{Pettini04} method or the \citet{Kewley02a} method to minimize any residual discrepancies, and to maintain accurate relative metallicities compared to other calibrations.  These two diagnostics have the added benefit that at high metallicities, the Kewley \& Dopita \NIIOII\ and Pettini \& Pagel \NIIHa\ calibrations are relatively independent of the method used to remove AGN.

Future work into tailored photoionization models, high S/N spectroscopy of luminous stars in the Milky Way and nearby galaxies, metal recombination lines, IR fine structure lines, and temperature fluctuation studies may help resolve the metallicity discrepancy problem in the near future.  Until then, only relative metallicity comparisons are reliable.

\acknowledgments
We thank Rolf Jansen for providing the B-R colors for the NFGS.  We thank Christy Tremonti  for useful discussions.   L. J. Kewley is supported by a Hubble Fellowship.   S. Ellison acknowledges an NSERC Discovery Grant which funded this research and is grateful to the IfA for hosting visits during which some of this work was completed.  L. J. Kewley is grateful to the Aspen Center for Physics, where some of this work was done.

\appendix

\section{Metallicity Calibrations: Equations and Method} \label{App_calibrations}

\subsection{Breaking the \R23\ Degeneracy} \label{R23_break}
    
Many empirical and theoretical metallicity calibrations rely on the \ratioR23\ line ratio, known as ``\R23".  The major drawback to using \R23\ is that it is double-valued with metallicity; \R23\ gives both a low metallicity estimate (``lower branch") and a high estimate (``upper branch") for most values of \R23\ \citep[see e.g.,][for a discussion]{Kobulnicky04}.  Additional line ratios, such as \NIIHa, or \NIIOII, are required to break this degeneracy.  

The SDSS catalog contains very few metal-poor galaxies \citep{Izotov04,Kniazev03,Kniazev04,Papaderos06a,Izotov06b}.    Metal poor galaxies are often lacking in magnitude-limited emission-line surveys because they are intrinsically rare, compact and faint \citep[e.g.,][]{Terlevich91,Masegosa94,VanZee00}.
For the purpose of investigating the upper and lower \R23\ branches, we supplement the SDSS sample with (a) the low metallicity galaxy sample described in \citet{Kewley07} and \citet{Brown06}, and (b) the 
 \citet{Kong02a} blue compact galaxy sample.

Note that we do not calculate an initial metallicity from an \NIIHa\ or \NIIOII\ metallicity calibration because in some cases, a systematic discrepancy between a metallicity calibration based on \NIIHa\ or \NIIOII\  and the calibration based on  \R23\ will cause galaxies to be improperly placed on the upper or lower branch of \R23.   For example, an \NIIHa\ metallicity calibration that systematically produces higher estimates than the subsequent \R23\ calibration may cause metallicities to be erroneously estimated from the upper \R23\ branch.  

\begin{figure}[!h]
\plotone{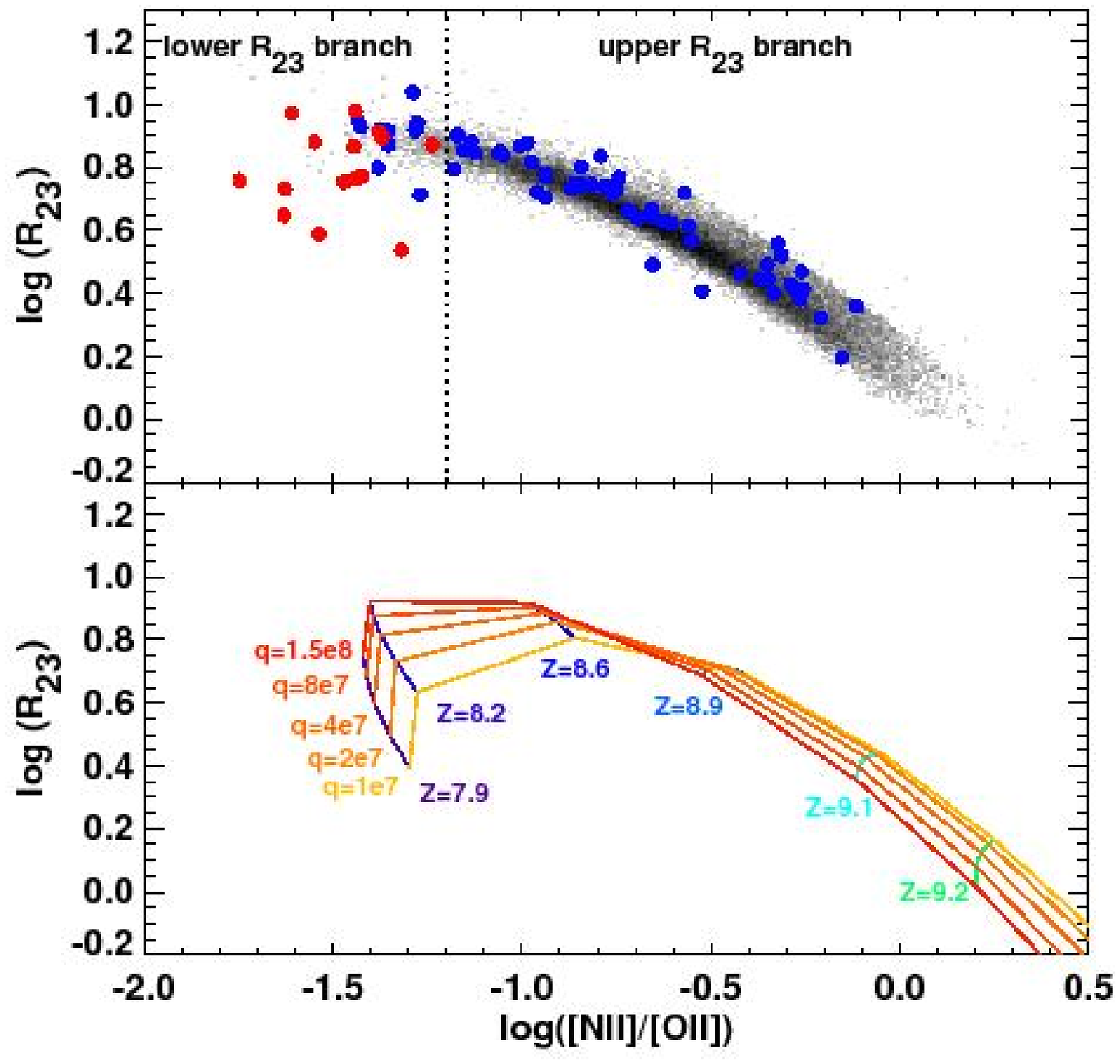} 
\caption{(a) The observed relationship between the metallicity-sensitive \ratioNIIOII\ line ratio and the 
commonly-used \ratioR23\ ratio.  The SDSS galaxies (black), the \citet{Kong02a} blue compact galaxy sample 
(blue) and the  \citet{Brown06} low metallicity galaxy samples (red) are shown. The [NII]/[OII] ratio is a strong monotonic function of metallicity to log([NII]/[OII])$\gtrsim -1.2$, while \R23\ has a maximum at log(\R23 )$\sim 0.9$.  For our samples, the \R23\ maximum is likely to occur at log([NII]/[OII])$\sim -1.2$.  This value can be used to
break the \R23\ degeneracy for galaxies where [NII]/[OII] can be corrected for extinction using the Balmer Decrement.   (b) The theoretical relationship between the [NII]/[OII] and \R23\ line ratios using the stellar population synthesis and photoionization model grids of Kewley \& Dopita (2002).  Models are shown for constant metallicities of \OH$=7.9, 8.2, 8.6, 8.9, 9.1, 9.2$ and ionization parameters of 
$q=1\times 10^7,\, 2\times10^7,\,4\times10^7,\,8\times 10^7,1.5\times 10^8$~$cm\, s^{-1}$.  We choose a break between the \R23\ upper and lower branches at  log([NII]/[OII])$\sim -1.2$, which corresponds to a metallicity of \OH$\sim 8.4$ according to the theoretical models.
\label{NIIOII_R23}}
\end{figure}

We use the \NIIOII\ ratio to break the \R23\ degeneracy for our SDSS sample.  The \NIIOII\ ratio is not sensitive to ionization parameter to within ($\pm 0.05$~dex), and it is a strong function of metallicity above log(\NIIOII)$\gtrsim -1.2$ \citep[][]{Kewley02a}.     Figure~\ref{NIIOII_R23}a shows that the division between the \R23\ upper and lower branches occurs at log(\NIIOII)$\sim -1.2$ for the SDSS and supplementary samples.    For comparison, Figure~\ref{NIIOII_R23}b shows the theoretical relationship between \NIIOII\ and \R23 using the population synthesis and photoionization models of  \citet{Kewley02a}.  The observed \R23\ peak at log(\NIIOII)$\sim -1.2$ corresponds to a metallicity of \OH$\sim 8.4$ according to the theoretical models.  

\placefigure{NIIOII_R23}
\epsscale{0.8}

\begin{figure}[!h]
\plotone{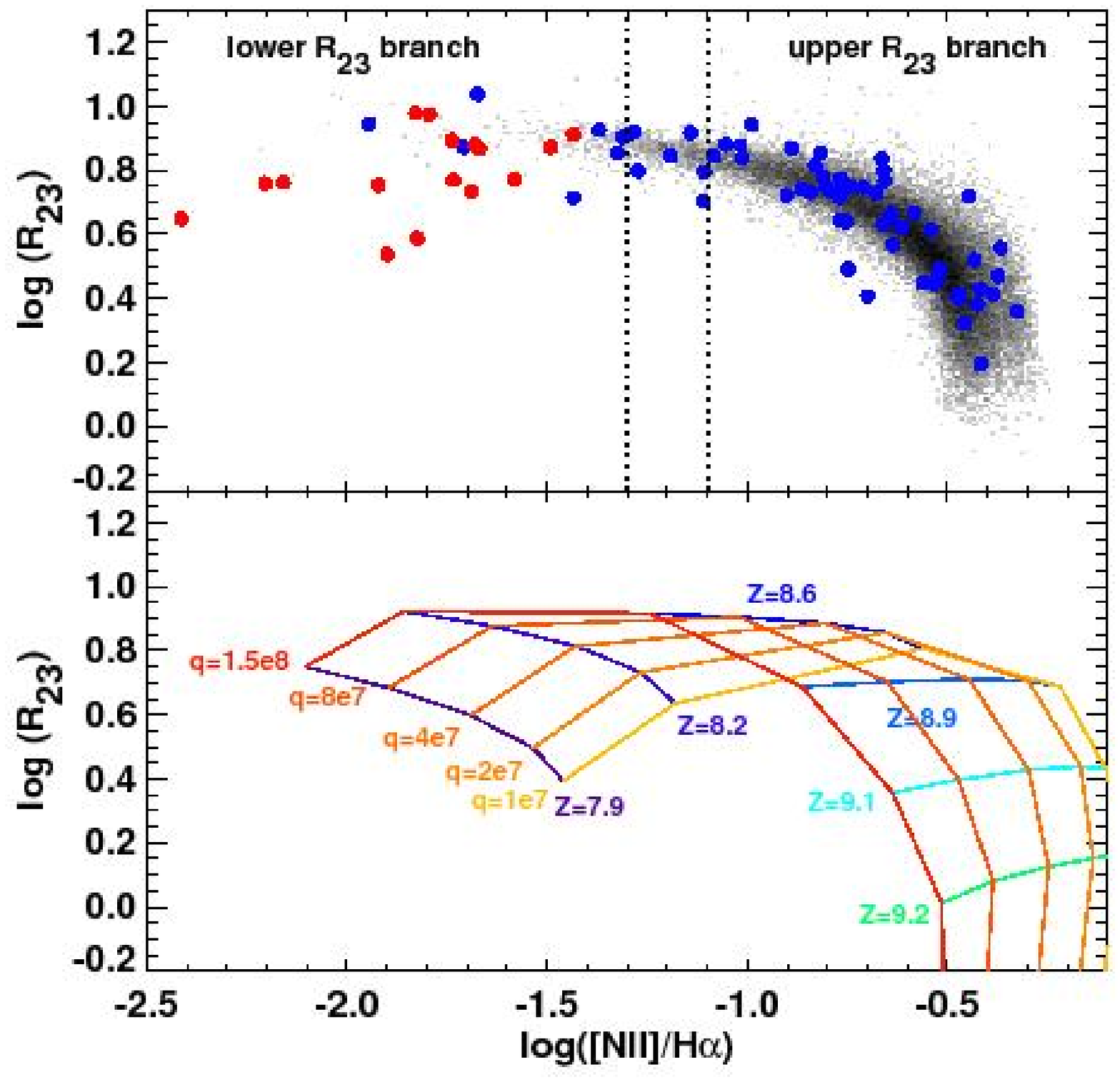} 
\caption{(a) The observed relationship between the metallicity-sensitive \ratioNIIHa\ line ratio and the  commonly-used \ratioR23\ ratio. The SDSS galaxies (black), the \citet{Kong02a} blue compact galaxy sample 
(blue) and the  \citet{Brown06} low metallicity galaxy samples (red) are shown.  For  log([NII]/H$\alpha$) $\gtrsim -1.1$, galaxies are likely to lie on the upper \R23\ branch.  For $-1.3<$log([NII]/H$\alpha$ )$< -1.1$, [NII]/H$\alpha$ cannot discriminate between the upper and lower \R23\ branches.  For log(\NIIHa )$\lesssim -1.3$, galaxies lie on the lower \R23\ branch.     (b) The theoretical relationship between the \NIIOII\ and \R23\ line ratios using the stellar population synthesis and photoionization model grids of Kewley \& Dopita (2002).  Models are shown for constant metallicities of \OH$=7.9,8.2,8.6,8.9,9.1,9.2$ and ionization parameters of  $q=1\times 10^7,\, 2\times10^7,\,4\times10^7,\,8\times 10^7,1.5\times 10^8$~$cm\, s^{-1}$.
\label{NIIHa_R23}}
\end{figure}

For galaxies at high redshift, the \NIIOII\ ratio cannot be used to break the \R23\ degeneracy because either (a) \NIIOII\ cannot be corrected for extinction due to a lack of reliable Balmer line ratios, and/or (b) \NII\ and \OII\ are not observed simultaneously in a given spectrum.  In this case the \NIIHa\ ratio is used (Figure~\ref{NIIHa_R23}).  Figure~\ref{NIIHa_R23}a shows that the division between the \R23\ upper and lower branches occurs between  $-1.3<$log(\NIIHa)$\lesssim -1.1$ for the SDSS and supplementary samples.   The division between the upper and lower \R23\ branches using \NIIHa\ (Figure~\ref{NIIHa_R23}a) is less clear than for \NIIOII\ (Figure~\ref{NIIOII_R23}a) because the \NIIHa\ ratio is less sensitive to metallicity, and more sensitive to ionization parameter, than \NIIOII.   

We check whether our empirical \NIIHa\ division between the upper and lower \R23\ branches $-1.3<$log(\NIIHa)$\lesssim -1.1$ is compatable with our \NIIOII\ division (log(\NIIOII)$\lesssim -1.2$) by comparing the number of galaxies placed on the upper and lower branches using each ratio.
For log(\NIIHa)$\lesssim -1.3$, the majority of galaxies (150/175; 86\%)  lie on the lower \R23\ branch according to their \NIIOII\ line ratios.    For  log(\NIIHa)$\gtrsim -1.1$, the upper \R23\ branch can be clearly seen.  For $-1.3<$log(\NIIHa)$<-1.1$, \NIIHa\ cannot discriminate between the upper and lower \R23\ branches.  Figure~\ref{NIIHa_R23}b shows that galaxies with $-1.3<$log(\NIIHa)$<-1.1$ are likely to have metallicities that are close to the \R23\ maximum.  For the SDSS galaxies with $-1.3<$log(\NIIHa)$<-1.1$ in Figure~\ref{NIIHa_R23}a, the \NIIOII\ ratios indicate that 634/1127 (56\%) of SDSS galaxies lie on the upper branch and 493/1127 (44\%) SDSS galaxies lie on the lower branch.  

For the \R23\ methods in this paper, we use the \NIIOII\ line ratio to break the \R23\ degeneracy.  

\placefigure{NIIHa_R23}

\subsection{Theoretical Photoionization Methods}

\subsubsection{McGaugh (1991) - M91} \label{M91}

The \citet{McGaugh91} calibration of \R23\ is based on detailed \HII\ region models using the photoionization code CLOUDY \citep{Ferland98}. The M91 calibration includes a correction for ionization parameter variations.  We use the \NIIOII\ line ratio to break the \R23\ degeneracy, as described in Section~\ref{R23_break}, and we apply the analytic expressions for the M91 lower and upper branches given in \citet{Kobulnicky99a}:

\begin{eqnarray}
12 + \log({\rm O/H})_{lower} & =  & 12-4.944 + 0.767 x +0.602 x^2 \nonumber \\
                                                   &   & - y ( 0.29 + 0.332 x -0.331 x^2)\label{M91_lower} 
\end{eqnarray}

\begin{eqnarray}
12 + \log({\rm O/H})_{upper} & = & 12 - 2.939 - 0.2 x -0.237 x^2  \nonumber \\
                                                   &     & - 0.305 x^3 - 0.0283 x^4 - y (0.0047 \nonumber \\
                                                   &   & -0.0221 x -0.102 x^2 -0.0817 x^3 \nonumber \\   
                                                   &    & -0.00717 x^4) \label{M91_upper}                                               
 \end{eqnarray}

where $x=\log ( {\rm R_{23}}) = \log \left(\frac{ {\rm [OII]} \lambda 3727 +  {\rm [OIII]} \lambda 4959+  {\rm [OIII]} \lambda 5007}{{\rm H} \beta}\right)$, and $y= \log ( {\rm O_{32}}) = \log \left( \frac{{\rm [OIII] \lambda 4959}+{\rm [OIII] \lambda 5007}}{{\rm [OII]} \lambda 3727}\right)$.  The estimated accuracy of the M91 calibration is $\sim 0.15$~dex.   

\subsubsection{Kewley \& Dopita (2002) - KD02} \label{KD02}

The \citet[][]{Kewley02a} calibrations are based on a self-consistent combination of detailed stellar population synthesis and photoionization models.    The estimated accuracy of the KD02 calibrations is $\sim 0.1$~dex.  This estimate is derived by varying the major assumptions in the stellar evolution and photoionization models (including the star formation prescription, electron density, and the initial mass function).   
KD02 outlined a ``recommended"
approach to deriving metallicities that uses the \NIIOII\ line ratio for high metallicities and a combination of \R23\ calibrations for lower metallicities.  We use a revised version of the KD02 calibration.  For 
log(\NIIOII)$> -1.2$  , we use the original KD02 \NIIOII\ metallicity calibration given by 
  
  \begin{equation}
 \log([NII]/[OII])= 1106.8660 - 532.15451Z + 96.373260 Z^2 -7.8106123 Z^3 + 0.23928247 Z^4 \label{KD02_NIIOII_Z}
  \end{equation}
   
 where $Z=\log(O/H)+12$.  We use the IDL task {\it fz\_roots} to solve the 4th order polynomial for $Z$.
The coefficients in Equation~\ref{KD02_NIIOII_Z} are based on the theoretical $q=2\times10^7$~cm/s 
relationship between \NIIOII and $Z$.  However, the detailed relationship between \NIIOII and $Z$ is independent of ionization parameter to within $\sim 0.1$~dex for log(\NIIOII)$> -1.2$ and the ionization parameters covered by the SDSS ($q=1\times 10^7 - 8\times 10^7$~cm/s).
   
For log(\NIIOII)$< -1.2$ (or KD02 \OH$<8.4$), KD02 recommend using an average of \R23\ methods.  In this regime, we use the average of the KK04 lower banch \R23\ calibration (equation~\ref{KK04_Z}) and the lower branch M91 \R23\ calibration (equation~\ref{M91_lower}).   Both of these calibrations correct 
for ionization parameter variations.

\subsubsection{Kobulnicky \& Kewley (2004) - KK04}  \label{KK04}

\citet{Kobulnicky04} use the stellar evolution and photoionization grids from \citet{Kewley02a} to produce  an improved fit to the \R23\ calibration.  The estimated accuracy of the KK04 method is $\sim 0.15$~dex.

The \R23\ calibration is sensitive to the ionization state of the gas, particularly for low metallicities where the \R23\ line ratio is not a strong function of metallicity.  The ionization state of the gas is characterized by the ionization parameter, defined as the number of hydrogen ionizing photons passing through a unit area per second, divided by the hydrogen density of the gas.  The ionization parameter $q$ has units of cm/s and can be thought of 
as the maximum velocity ionization front that a radiation field is able to drive through the nebula.   The 
ionization parameter is typically derived using the \OIII/\OII\ line ratio.  This ratio is sensitive to metallicity and therefore KK04 recommend an iterative approach to derive a consistent ionization parameter and metallicity solution.   We first use the \NIIOII\ ratio to determine whether each SDSS galaxy lies on the 
upper or lower \R23\ branch.  We then calculate an initial ionization parameter by assuming a nominal lower branch (\OH$=8.2$) or upper branch (\OH$=8.7$) metallicity using equation (13) from KK04, i.e.

\begin{eqnarray}
\log(q) & = & \{32.81 - 1.153 y^2 \label{KK04_q} \nonumber \\
            &   & + [12 + \log({\rm O/H})](-3.396-0.025y+0.1444y^2)\} \nonumber \\
            &    &\times \{ 4.603 - 0.3119 y -0.163 y^2 \nonumber \\
            &    & +  [12 + \log({\rm O/H})](-0.48 +0.0271 y + 0.02037 y^2 )\}^{-1} 
\end{eqnarray}

where $y=\log O_{32} = \log \left({[{\rm OIII}]\,\lambda 5007}/{[{\rm OII}]\,\lambda 3727}\right)$.  The initial resulting ionization parameter is used to derive an initial metallicity estimate from KK04 equation (16) for log(\NIIOII)$<-1.2$ (\OH$\lesssim8.4$), or KK04 equation (17) for log(\NIIOII)$>-1.2$ (\OH$\gtrsim8.4$):

\begin{eqnarray}
12 + \log({\rm O/H})_{lower} & = &  9.40  + 4.65 x  - 3.17 x^2 \nonumber  \\
              & &   - \log(q)(0.272 +0.547 x -0.513 x^2) \label{KK04_Z_up}
\end{eqnarray}

\begin{eqnarray}
12 + \log({\rm O/H})_{upper} & = &  9.72  - 0.777 x - 0.951 x^2 -0.072 x^3  -0.811 x^4  x  \nonumber  \\
              & &   - \log(q)(0.0737 -0.0713 x -0.141 x^2 + 0.0373 x^3 - 0.058 x^4) \label{KK04_Z}
\end{eqnarray}

where $x = \log {\rm R}_{23} = \log \left( \frac{[{\rm OII}]\, \lambda 3727 + [{\rm OIII}]\,\lambda \lambda 4959,5007}{H \beta} \right)$.
Equations~\ref{KK04_q} and \ref{KK04_Z} (or \ref{KK04_Z_up}) are iterated until \OH\ converges.
Three iterations are typically required to reach convergence.

\subsubsection{Zaritsky et al. (1994) - Z94} \label{Z94}

The \citet{Zaritsky94} calibration is based on the \R23\ line ratio.  This calibration is derived from the average of three previous calibrations by \citet{Edmunds84,Dopita86,McCall85}.
The uncertainty in the Z94 calibration is estimated by the difference in metallicity estimates between the three calibrations.   Z94 provide a polynomial fit to their calibration that is only valid for the upper \R23\ branch (i.e. \OH$>8.4$, or log(\NIIOII)$>-1.2$).

\begin{equation}
12+\log({\rm O/H}) = 9.265 - 0.33 x -0.202 x^2 -0.207 x^3 -0.333 x^4 \label{Z94_Z}
\end{equation} 

where $x = \log {\rm R}_{23} = \log \left( \frac{[{\rm OII}]\, \lambda 3727 + [{\rm OIII}]\,\lambda \lambda 4959,5007}{H \beta} \right)$.   A solution for the ionization parameter is not explicitly included in the Z94 calibration.

\subsubsection{Tremonti et al. (2004) - T04}\label{T04}

T04 estimated the metallicity for each galaxy statistically based on theoretical model fits to the strong emission-lines \OII, \Hb, \OIII, \Ha, \NII, \SII.  The model fits were calculated using a combination of stellar population synthesis models from \citet{Bruzual03} and CLOUDY photoionization models \citet{Ferland98}.    The T04 scheme is more sophisticated than the other theoretical methods because it takes advantage of all of the strong  emission lines in the optical spectrum, allowing more constraints to be made on the model parameters.    Calibrations of various line ratios to the theoretical T04 method are given by \citet{Nagao06} and  \citet{Liang06}.
We use the original T04 metallicities, available from {\it http://www.mpa-garching.mpg.de/SDSS/}  for this study.

\subsection{\Te\ method}\label{Te}

We derive the gas-phase oxygen abundance  following the procedure
outlined in \citet{Izotov06a}.  This procedure utilizes the
electron-temperature (\Te) calibrations of \citet{Aller84} and the atomic data
compiled by \citet{Stasinska05}.   Abundances are determined within the framework of 
the classical two-zone HII-region model \citep{Stasinska80}.
The ratio of the auroral \OIII~$\lambda 4363$
and \OIII~$\lambda \lambda 4959,5007$ emission-lines gives an electron
temperature in the O$^{++}$ zone.   We  derive electron densities measured using the 
\SII~$\lambda 6717$/\SII~$\lambda 6731$ line ratio.
These electron temperatures are insensitive
to small variations in electron density; we obtain the same ${\rm T_{e}}$ with an electron density of 367~cm$^{-3}$.  The electron temperature of the 
O$^{+}$ zone is calculated assuming ${\rm T_{e}}({\rm O}^{+})=0.7\, {\rm T_{e}}({\rm O}^{++} ) + 0.3 $
\citep{Stasinska80}.  We calculate the metallicity in
the O$^{+}$ and O$^{++}$ zones assuming
	\begin{equation}
{\rm O/H}={\rm O}^{+}/{\rm H}^{+} + {\rm O}^{++}/{\rm H}^{+}
\end{equation}

The uncertainty in the absolute O/H metallicity determination by this \Te\ method
is $\sim 0.1$~dex. This intrinsic uncertainty is the dominant error in our \Te\ metallicity
determination, and includes errors in the use of simplified \HII\ region models and
possible problems with electron temperature fluctuations \citep{Pagel97}.  Fortunately, these errors 
affect all \Te-based methods in a similar way and the error in relative metallicities derived using the same method is likely to be $<< 0.1$~dex.   

This ``classical" \Te\ approach does not take unseen stages of ionization or electron temperature fluctuations into account.   \citet{Bresolin06} notes that if electron temperature fluctuations are substantial and are not taken into account, \Te-based calibrations can only provide a lower limit to the true metallicity, particularly in the high metallicity regime where \Te\ fluctuations appear stronger.  
We find that the \Te\ method does not produce any SDSS metallicities of solar \citep[\OH$\sim 8.69$; ][]{Allende01,Asplund04} or above, even for galaxies where the fiber only captures $20-30$\% of the central g'-band galaxy light.   Covering fractions of $20-30$\% correspond to diameters of $\sim 5$~kpc for the mean size of nearby star-forming galaxies \citep{Kewley05a}.  Spiral galaxies typically have metallicities within similar apertures that are $\sim 1-2\times$solar, measured using various independent methods \citep[see][for a review]{Henry99}. For example, our Galaxy has consistent central metallicities within the central $\sim 5$~kpc of $1-2\times$solar from studies of planetary nebulae \citep{Martins00}, IR fine structure lines in \HII\ regions \citep{Simpson95,Afflerbach97}, and radio recombination lines \citep{Quireza06}.   

We conclude that \Te\ metallicities should be used with caution when other line ratios (such as \NIIHa\ and \NIIOII) indicate upper branch (Figures~\ref{NIIOII_R23} and~\ref{NIIHa_R23}) or supersolar metallicities.

\subsection{Empirical \Te\ fit methods}

\HII\ regions with electron temperature-based metallicities have been used in many studies to derive
empirical fits to strong-line ratios that can be applied to \HII\ regions and galaxies where the \O4363\ line is not observed.   

\subsubsection{Pettini \& Pagel (2004) - PP04 O3N2 \& N2 }\label{PP04}

\citet{Pettini04} derived two new methods for measuring metallicities in galaxies at high redshift.  At high redshift, obtaining a reddening estimate is difficult and in some cases, impossible, and flux calibration in the infrared is non-trivial.  Ratios of lines that are very close in wavelength do not require reddening correction or flux calibration.  PP04 fit the observed relationships between \NIIHa,  (\OIIIHb)/(\NIIHa)  and the \Te-based metallicity for a sample of 137 \HII\ regions.  Of these 137 \HII\ regions, 131 have \Te-based metallicities and 6 high metallicity \HII\ regions have metallicities derived using detailed photoionization models.   Because the vast majority of \HII\ regions in the PP04 sample have \Te-based metallicities, we refer to the PP04 method as an empirical \Te\ fit method.    The fit to the relationship between \Te\ metallicities and the (\OIIIHb)/(\NIIHa) ratio is:

\begin{equation}
12 + \log({\rm O/H}) = 8.73 - 0.32 \times {\rm O3N2} \label{PP04_O3N2_Z},
\end{equation}

where O3N2 is defined as ${\rm O3N2} = \log \left( \frac{[OIII]\,\lambda 5007 / H\beta}{[NII]\, \lambda 6584 / H \alpha} \right)$.  Equation~\ref{PP04_O3N2_Z} is only valid for O3N2$\gtrsim 2$.   We refer to this calibration as "PP04 O3N2".

PP04 fit the relationship between \Te\ metallicities and the \NIIHa\ ratio by a line and a third-order polynomial.  We use the polynomial fit given by

\begin{equation}
12 + \log({\rm O/H}) = 9.37 + 2.03  \times {\rm N2} + 1.26 \times {\rm N2}^2 + 0.32 \times {\rm N2}^3, \label{PP04_N2_Z}
\end{equation}

where N2 is defined as ${\rm N2} = \log ([NII]\, \lambda 6584 / H \alpha)$.  Equation~\ref{PP04_N2_Z} is valid for $-2.5<{\rm N2}<-0.3$.  

\begin{figure}[!t]
\plotone{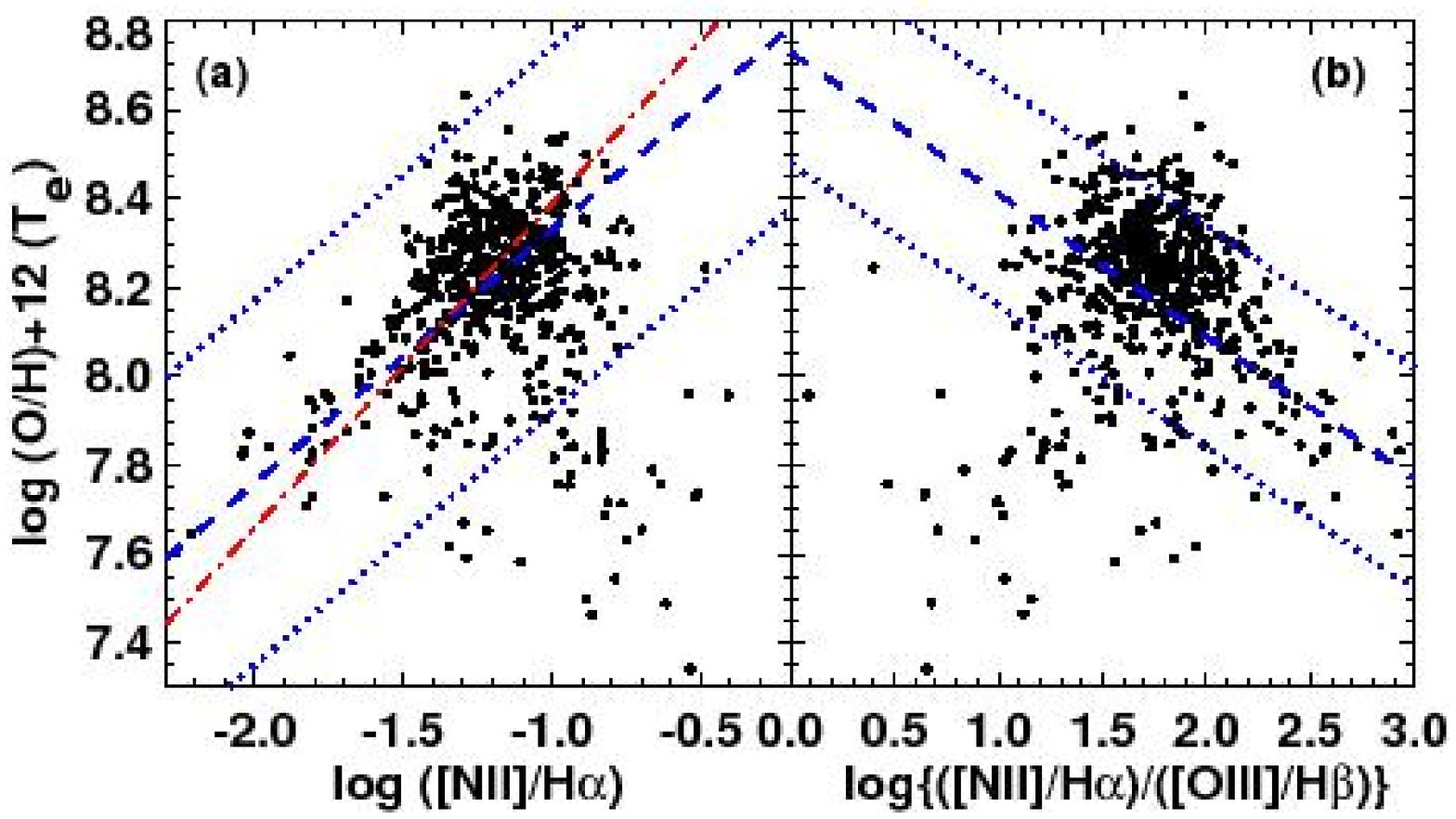} 
\caption{The observed relationship between the metallicities derived using the \Te\  method and (a) the \ratioNIIHa\ line ratio, and (b) the  \OIIIHb/\NIIHa\  ratio.  SDSS galaxies with useable \OIII~$\lambda 4363$ line fluxes (S/N$>3$) are shown as black filled circles.  The PP04 calibration (dashed lines) and 95 percentile lines (dotted lines) are shown for each line ratio.   The D02 calibration (red dot-dashed line) is shown for panel (a).  A fraction of SDSS-\Te\ galaxies have \Te\ metallicities that lie below the 95 percentile line in both the [NII]/H$\alpha$ and \OIIIHb/\NIIHa\ diagrams. These galaxies have high [NII]/H$\alpha$ and [NII]/[OII] ratios, suggesting high (above solar) metallicities.  The \Te-method appears to underestimate the metallicity in these galaxies, possibly as a result of temperature gradients or fluctuations that may occur preferentially at high metallicities.
\label{PP04_SDSS}}
\end{figure}

Because the PP04 method was derived using a fit to \HII\ regions rather than galaxies, we check whether the PP04 relations are suitable for metallicity estimates of the SDSS sample.
In Figure~\ref{PP04_SDSS} we show the relationship between (a) N2 and \Te\ metallicities, and 
(b) O3N2  and \Te\ metallicities for the SDSS galaxies in our sample with measurable (S/N$>3$) \OIII~$\lambda 4363$ lines.    The dashed line indicates the PP04 calibrations based on \HII\ regions, while the dotted lines encompass 95\% of the \HII\ regions in the PP04 sample.  The majority of the SDSS galaxies lie within the PP04 95 percentile lines.  However, 47/546 (9\%) and 69/546 (13\%) of SDSS-\Te\ galaxies have \Te\ metallicities that lie below the 95 percentile line in the N2 and N2O3 diagrams, respectively.  These galaxies have high \NIIHa\ and \NIIOII\ ratios (log(\NIIHa)$>-1$; log(\NIIOII)$>-0.8$), indicating supersolar metallicities, according to all of the \NIIHa\ and \NIIOII-based metallicity diagnostics.   Both Figure~\ref{PP04_SDSS} and the high \NIIHa\ and \NIIOII\ ratios suggest that  the \Te-method underestimates metallicities for galaxies that lie below the PP04 95 percentile line.   
 
\placefigure{PP04_SDSS}

\subsubsection{Pilyugin (2005) -  P05}\label{P05}

\citet{Pilyugin01} derived an empirical calibration for \R23\ based on \Te-metallicities for a sample of \HII\ regions.  This calibration has been updated by \citet[][; hereafter P05]{Pilyugin05}, using a larger sample of \HII\ regions.  They perform fits to the relationship between \R23\ and \Te-metallicities that includes an excitation parameter $P$ that corrects for the effect of ionization parameter.  The resulting calibration has an 
upper branch calibration that is valid for \Te-based metallicities \OH$>8.25$, and a lower branch 
calibration that is valid for \Te-based metallicities \OH$<8.0$.    We use the \NIIOII\ ratio (Figure~\ref{NIIOII_R23}) to discriminate between the upper and lower branches for P05, and we apply the appropriate upper and lower-branch calibrations (equations~22 and 24 in P05):

\begin{equation}
12+\log({\rm O/H})_{upper} = \frac{R_{23} + 726.1 + 842.2 P + 337.5 P^2}{85.96 + 82.76 P + 43.98 P^2 + 1.793 R_{23}}
\end{equation}

\begin{equation}
12+\log({\rm O/H})_{lower} = \frac{R_{23} + 106.4 + 106.8 P - 3.40 P^2}{17.72 + 6.60 P + 6.95 P^2 - 0.302 R_{23}}
\end{equation}

where $R_{23} = \frac{[{\rm OII}]\, \lambda \lambda 3727,29 + [{\rm OIII}]\, \lambda \lambda 4959,5007}{{\rm H} \beta}$, and $P= \frac{ [{\rm OIII}]\, \lambda \lambda 4959,5007 / {\rm H}\beta}{R_{23}}$ .
P05 estimate that the accuracy for reproducing \Te-based metallicities with the P05 calibration is $\sim 0.1$~dex 

\begin{figure}[!t]
\plotone{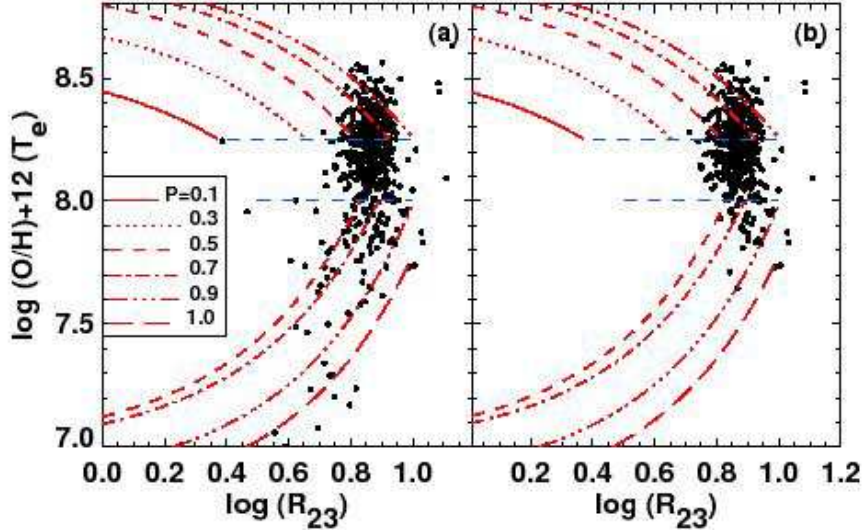} 
\caption{The observed relationship between the metallicities derived using the \Te\  method and the \R23\ line ratio for (a) all SDSS galaxies in our sample with measurable (S/N$>3$) \OIII~$\lambda 4363$ fluxes, and (b) for the SDSS galaxies in our sample with measurable \OIII~$\lambda 4363$ lines that lie above the lowest 95 percentile line in the PP04 O3N2 calibration (Figure~\ref{PP04_SDSS}).
\label{Te_P05}}
\end{figure}

Because the P05 method was derived using fits to \HII\ regions, we test whether the P05 method is applicable to the relationship between the SDSS \Te\ metallicities and \R23\ in Figure~\ref{Te_P05}.   In  Figure~\ref{Te_P05}a, we plot all SDSS galaxies in our sample with S/N ratio $>3$ in the \OIII~$\lambda 4363$ line.  The upper and lower P05 branches are shown for different values of the excitation parameter $P$ (red dot-dashed lines).  Several galaxies lie outside the bounds of the P05 lower branch (\OH\ (Te)$<8.0$).

In Figure~\ref{Te_P05}b, we exclude the galaxies that lie below the lowest 95 percentile line in the PP04 O3N2 calibration (Figure~\ref{PP04_SDSS}b) that are likely to have unreliable \Te\ metallicities.   As we discussed in Section~\ref{PP04}, these excluded galaxies have  \NIIHa\ and \NIIOII\ ratios that indicate metallicities above solar.   We note that the excluded galaxies
have excitation parameters between $0.2<P<0.8$, with a mean excitation parameter of $0.46\pm0.03$.  These $P$ values are
more consistent with the P05 upper branch (range $0.2<{\rm P}<0.8$, mean $0.64 \pm 0.03$) than the P05 lower branch 
(range $0.6<{\rm P}<1.0$, mean $0.8\pm0.1$) for our SDSS sample.   The \Te\ method may not be reliable for these galaxies.

\placefigure{Te_P05}

\subsection{Combined \Te-strong-line method}

\subsubsection{Denicolo, Terlevich \& Terlevich (2002) - D02}

The \citet{Denicolo02} calibration is based on a fit to the relationship between the \Te\ metallicities and the \NIIHa\ line ratio for $\sim 155$ \HII\ regions.  Of these \HII\ regions, $\sim 100$ have metallicities derived using the \Te\ method, and $55$ \HII\ have metallicities estimated using the theoretical M91 \R23\ method, or an empirical method proposed by \citet{Diaz00} method based on the sulfur lines.    The division between \HII\ regions with \Te-based metallicities and those with strong-line metallicities occurs at \OH$\sim 8.4$.  The D02 calibration is given by a linear least-squares fit:

\begin{equation}
12+\log({\rm O/H}) = 9.12 + 0.73 \times N2
\end{equation}

where $N2 = \log( [{\rm NII}]\,\lambda 6584 / {\rm H}\alpha )$.  D02 estimate that the uncertainty the derived metallicities is $\sim 0.2$~dex.  

In Figure~\ref{PP04_SDSS}, we compare the D02 fit (red dot-dashed line) to the \NIIHa-\Te relationship for the 
SDSS galaxies.  The D02 method provides a reasonable fit to the SDSS galaxies, given the large scatter, and is similar to the PP04 N2 curve to within $\sim 0.2$~dex over the metallicity range $7.4<12+\log({\rm O/H})<8.8$.

\section{Metallicity Conversions: Worked Examples} \label{worked_examples}

Three galaxies have metallicities of \OH$=8.3, 8.6$, and $9.1$ calculated using three different methods; KK04, PP04, and D02, respectively.  To compare these galaxy metallicities with those derived by the SDSS team \citep{Tremonti04}, we convert the three galaxy metallicities into a metallicity base of T04.  

For a galaxy with metallicity \OH$_{KK04}=8.3$, the KK04 metallicity is calculated from the lower \R23\ branch (see Section~\ref{KK04}).  Table~\ref{conv_table} gives the coefficients of the polynomial that converts a KK04 metallicity into a T04 base metallicity.  Because our KK04 metallicity is from the lower branch, we use the linear relation for the lower branch conversion:

\begin{eqnarray}
[\log{\rm (O/H)}+12]_{T04} & =  & -4.5710 + 1.53261 \times [\log{\rm (O/H)}+12]_{KK04}\nonumber \\
 & \sim & 8.2 \nonumber
\end{eqnarray}

To convert an original PP04 metallicity (\OH$_{PP04}=8.6$) from O3N2 into a T04 base metallicity, Table~\ref{conv_table} gives

\begin{eqnarray}
[\log{\rm (O/H)}+12]_{T04} & = & -738.1193  + 258.96730 \times [\log{\rm (O/H)}+12]_{PP04} \nonumber \\
  &    & - 30.057050 \times [\log{\rm (O/H)}+12]^{2}_{PP04} \nonumber \\
  &     & +1.167937 \times [\log{\rm (O/H)}+12]^{3}_{PP04} \nonumber \\
 & \sim  & 8.9 \nonumber
\end{eqnarray}

An original D02 metallicity \OH$_{D02} =9.1$ cannot be converted into a T04 metallicity because this D02 abundance is above the valid range for our conversion from D02 into T04 ($8.05 - 8.9$).

As a final example, we convert a T04 abundance of \OH$_{T04}=8.3$ into an M91 base metallicity.  Note that the valid upper and lower branch ranges overlap for the conversion of T04 into M91, ie. $8.2 < \log({\rm O/H})_{T04} < 9.2$ (upper branch) and 
$8.05 < \log ({\rm O/H})_{T04}< 8.4$ (lower branch).  At a T04 metallicity between $8.2< \log({\rm O/H})_{T04} < 8.3$, the M91 
\R23\ calibration is reaching a local maximum and is insensitive to metallicity.   In this regime, the M91 upper or lower branch should be selected based on the \NIIOII\ or \NIIHa\ ratio (Section~\ref{R23_break}).  If the log(\NIIHa) ratio in our example galaxy is $-0.7$, then Figure~\ref{NIIHa_R23} indicates that the metallicity is on the \R23\ upper branch, and therefore, the conversion from a T04 metallicity of \OH$_{T04}=8.3$ into an M91 base metallicity is 

\begin{eqnarray}
[\log{\rm (O/H)}+12]_{M91} & = & 404.1716   - 131.53250 \times [\log{\rm (O/H)}+12]_{T04} \nonumber \\
  &    & +14.49175 \times [\log{\rm (O/H)}+12]^{2}_{T04} \nonumber \\
   &     &  -0.5285842 \times [\log{\rm (O/H)}+12]^{3}_{T04} \nonumber \\
 & \sim  & 8.6 \nonumber
\end{eqnarray}

\bibliography{library}

\clearpage
\begin{deluxetable}{rllll}
\tablecolumns{5}
\tabletypesize{\scriptsize}
\tablecaption{Comparison of the 10 metallicity calibrations \label{calib_table}}
\tablehead{\# & ID & Emission lines & Calibration Class & Reference}
  \startdata
1 & T04\tablenotemark{a} & [OII], \Hb, [OIII], \Ha, [NII], [SII]\ & theoretical & \citet{Tremonti04} \\
2 & Z94  & \R23  & theoretical &  \citet{Zaritsky94} \\
3 & KK04 & \R23, [OIII]/[OII] & theoretical & \citet{Kobulnicky04} \\
4 & KD02 & [NII]/[OII], \R23, [OIII]/[OII] & theoretical & \citet{Kewley02a} \\
5 & M91 & \R23, [OIII]/[OII] & theoretical & \citet{McGaugh91} \\
6 &  D02 & [NII]/\Ha & combined  &  \citet{Denicolo02} \\
7 & PP04 & [NII]/Ha, [OIII]/\Hb & empirical & \citet{Pettini04} \\
8 & PP04 & [NII]/Ha & empirical & \citet{Pettini04} \\
9 & P01,P05 & \R23, [OIII]/[OII] & empirical & \citet{Pilyugin01,Pilyugin05} \\
10 & T$_{e}$ & [OIII]~$\lambda4363$, [OIII]~$\lambda \lambda4959,5007$ & direct & \citet{Aller84,Stasinska05,Izotov06b}
\enddata
\tablenotetext{a}{The T04 method uses a statistical technique to calculate the probability distribution 
of an object having a particular metallicity based on model fits to the  [OII], \Hb, [OIII], \Ha, [NII], [SII] emission lines.}
\end{deluxetable}

\begin{deluxetable}{llllll}
\tablecolumns{6}
\tabletypesize{\scriptsize}
\tablecaption{Robust fits\tablenotemark{a} to the MZ relations for the 9 strong-line metallicity calibrations \label{MZ_table}}
\tablehead{ ID & a & b & c & d & rms residuals}
\startdata
T04 &     -0.694114 &       1.30207 &    0.00271531 &   -0.00364112 &      0.12 \\
Z94 &       72.0142 &      -20.6826 &       2.22124 &    -0.0783089 &      0.13 \\
KK04 &       27.7911 &      -6.94493 &      0.808097 &    -0.0301508 &     0.10 \\
KD02 &       28.0974 &      -7.23631 &      0.850344 &    -0.0318315 &      0.10 \\
M91 &       45.5323 &      -12.2469 &       1.32882 &    -0.0471074 &      0.11 \\
D02 &      -8.71120 &       4.15003 &     -0.322156 &    0.00818179 &     0.08 \\
PP04 O3N2 &       32.1488 &      -8.51258 &      0.976384 &    -0.0359763 &     0.10 \\
PP04 N2 &       23.9049 &      -5.62784 &      0.645142 &    -0.0235065 &     0.09 \\
P01 &       91.6457 &      -25.9355 &       2.67172 &    -0.0909689 &      0.12 \\
P05 &       41.9463 &      -10.3253 &       1.04371 &    -0.0347747 &      0.13 \\
\enddata
\tablenotetext{a}{Robust fits are of the form $y=a+bx+bx^2+bx^3$ where $y=\log({\rm O/H})+12$ and $x=\log({\rm M})$ where M is the stellar mass in solar units}
\end{deluxetable}

\clearpage
        \LongTables 
        \begin{landscape} 
\begin{deluxetable}{lrrrrrrrr}
\tablecolumns{9}
\tablewidth{0pt}
\tabletypesize{\scriptsize}
\tablecaption{Metallicity Calibration Conversion Constants\label{conv_table}\tablenotemark{a}}
\tablehead{ID & KK04 & 	Z94 & 	M91 & 	KD02 & 	T04 & 	D02 & 	PP04 O3N2 & 	PP04 N2 \\
y & ({\it x})  & ({\it x})  & ({\it x})  & ({\it x})  & ({\it x})  & ({\it x})  & ({\it x})  & ({\it x}) }
\startdata
KK04  &  &  &  & & & & 	&  \\
x-range\tablenotemark{b} &  & 8.4-9.3  & 	 8.4-9.1;  8.0-8.25  & 	 8.4-9.2;  8.05-8.3   & 8.3-9.2  ;  8.05-8.4   & 8.2-8.9 ; 8.05-8.4   &  8.2-8.9 ;  8.05-8.3  & 	8.2-8.9 ;  8.05-8.3    \\ 
branch\tablenotemark{c} &  &  &	  up;  low  \phn \phd \phn \phd & 	 up;  low  \phn \phd \phn \phd & up;  low  \phn \phd \phn \phd & up;  low  \phn \phd \phn \phd & up;  low  \phn \phd \phn \phd & up;  low  \phn \phd \phn \phd \\ 
 {\it a} &  & 	348.1710 & 	-355.2968; 2.0021\phm{ ;} & 	1149.479; 1.0944\phm{ ;} & 	-162.1918; 5.0521\phm{ ;} & 	-1.7278; 6.5483\phm{ ;} & 	389.1568; 6.5339\phm{ ;} & 	211.1405; 5.2218\phm{ ;} \\
 {\it b} &  & 	-117.65370 & 	114.8835; 0.77696\phm{;}  & 	-389.9349; 0.87842\phm{;} & 	57.57935; 0.39887\phm{;} & 	1.52612; 0.21508\phm{;} & 	-133.79070; 0.21761\phm{;} & 	-81.11275; 0.37813\phm{;} \\
 {\it c} &  & 	13.502640 & 	-12.09838; \phm{000000;} & 	44.33080;   \phm{000000;} & 	-6.533486; \phm{000000;} & 	-0.034389; \phm{000000;} & 	15.59350; \phm{000000;} & 	10.577470;  \phm{000000;}\\
 {\it d} &  & 	-0.5130281 & 	0.4258375;  \phm{000000;}  & 	-1.675994;  \phm{000000;}   & 	0.249956;  \phm{000000;}   & 	;  \phm{000000;}   & 	-0.6020232;  \phm{000000;}   & 	-0.4510567; \phm{000000;}    \\
 $\rho_r$ &  & 	0.003 & 	0.007; 0.007 \phm{0;} & 	0.031; 0.004  \phm{0;} & 	0.048; 0.029  \phm{0;} & 	0.065; 0.039  \phm{0;} & 	0.041; 0.041 \phm{0;}  & 	0.064; 0.038  \phm{0;} \\ \hline
Z94 &  &  &  & & & & 	&  \\
x-range\tablenotemark{b} & 8.55-9.2 & 	  & 8.4-9.1  & 	 8.4-9.2  & 8.4-9.2  & 8.05-8.9  & 	 8.05-8.9  & 8.05-8.8   \\ 
 {\it a} & -1112.0910 & 	 & 	-868.280 & 	1086.903 & 	11.2595 & 	63.6386 & 	230.9335 & 	40.5515 \\
 {\it b} & 379.11370 & 	 & 	291.6262 & 	-366.5700 & 	-1.47641 & 	-13.87785 & 	-76.73906 & 	-8.67461 \\
 {\it c} & -42.880040 & 	 & 	-32.42779 & 	41.41521 & 	0.136681 & 	0.872216 & 	8.711059 & 	0.581691 \\
 {\it d} & 1.6218750 & 	 & 	1.206416 & 	-1.554630 & 	 & 	 & 	-0.3244087 & 	 \\
 $\rho_r$ & 0.005 & 	 & 	0.010 & 	0.040 & 	0.056 & 	0.081 & 	0.052 & 	0.076 \\ \hline
M91 &  &  &  & & & & 	&  \\
x-range\tablenotemark{b} & 8.25-9.15 & 	 8.4-9.2  & & 	 8.4-9.2 ;  8.1-8.4   & 8.2-9.2  ;  8.05-8.4   &  8.2-8.9 ;  8.05-8.4 & 8.2-8.9;  8.05-8.4 & 	 8.2-8.8 ; 8.05-8.3   \\ 
branch\tablenotemark{c} &  & 	  & & 	 up; low  \phn \phd \phn \phd& up; low  \phn \phd \phn \phd & up; low   \phn \phd \phn \phd & 8.2-8.9;  8.05-8.4 & up; low  \phn \phd \phn \phd \\ 
 {\it a} & -641.2458 & 	393.9855 & 	 & 	890.1334; -1.0907 & 	404.1716; 3.0825 \phm{ ;} & 	85.2839; 4.7927  \phm{ ;} & 	267.3936; 2.4196  \phm{ ;} & 	87.3710; 0.7641  \phm{ ;} \\
 {\it b} & 226.42050 & 	-127.86040 & 	 & 	-295.90760; 1.12114 & 	-131.53250; 0.61779 & 	-18.63342; 0.40796 & 	-85.20014; 0.70167 & 	-19.39959; 0.90362 \\
 {\it c} & -26.374200 & 	14.050330 & 	 & 	33.004210; \phm{000000;} & 	14.491750; \phm{000000;} & 	1.130870; \phm{000000;} & 	9.219665; \phm{000000;} & 	1.193544;  \phm{000000;} \\
 {\it d} & 1.0268860 & 	-0.5109532 & 	 & 	-1.2227730;  \phm{000000;} & 	-0.5285842;  \phm{000000;} & 	;  \phm{000000;} & 	-0.3267103;  \phm{000000;} & 	;  \phm{000000;} \\
 $\rho_r$ & 0.008 & 	0.008 & 	 & 	0.034; 0.004  \phm{0;} & 	0.047; 0.044  \phm{0;}& 	0.071; 0.059  \phm{0;}& 	0.048; 0.076  \phm{0;}& 	0.064; 0.070  \phm{0;} \\ \hline
KD02 &  &  &  & & & & 	&  \\
x-range\tablenotemark{b} & 8.2-9.2 & 	 8.4-9.2  & 8.5-9.1 ;  8.05-8.3   & 	& 8.2-9.2 ;  8.05-8.4   & 8.2-8.9 ;  8.05-8.4   & 	 8.2-8.9 ;  8.05-8.3   & 8.2-8.9 ; 8.05-8.3   \\ 
branch\tablenotemark{c} & &  & up; low   \phn \phd \phn \phd  &	& up; low   \phn \phd \phn \phd & up; low   \phn \phd \phn \phd & up; low  \phn \phd \phn \phd 	 &  up; low  \phn \phd \phn \phd 	\\ 
 {\it a} & 75.5327 & 	-476.8939 & 	-2127.7470; 1.0007  \phm{ ;} & 	  & 	387.2871; 4.1582  \phm{ ;} & 	47.3054; 5.9875  \phm{ ;} & 	159.0567; 5.8961  \phm{ ;} & 	1094.5410; 4.5323  \phm{ ;} \\
 {\it b} & -23.64323 & 	160.45270 & 	720.35980; 0.88853 & 	  & 	-129.69190; 0.49751 & 	-10.06952; 0.27338 & 	-54.18511; 0.28562 & 	-388.67530; 0.45232 \\
 {\it c} & 2.658484 & 	-17.752170 & 	-81.051380; \phm{000000;} & 	 \phm{000000;} & 	14.718030; \phm{000000;} & 	0.649502; \phm{000000;} & 	6.395364; \phm{000000;} & 	46.233760;  \phm{000000;} \\
 {\it d} & -0.0948578 & 	0.6579900 & 	3.0435570;  \phm{000000;} & 	  \phm{000000;} & 	-0.5531547;  \phm{000000;} & 	;  \phm{000000;} & 	-0.2471693;  \phm{000000;} & 	-1.8276270; \phm{000000;}  \\
 $\rho_r$ & 0.038 & 	0.036 & 	0.038; 0.003  \phm{0;} & 	 & 	0.041; 0.039  \phm{0;} & 	0.048; 0.055 \phm{0;}  & 	0.047; 0.059  \phm{0;} & 	0.047; 0.054 \phm{0;}  \\ \hline
T04 &  &  &  & & & & 	&  \\
x-range\tablenotemark{b} &  8.6-9.15 ;  8.2-8.4   & 8.4-9.2  & 8.6-9.1 ;  8.05-8.4   & 8.1-9.2 & 	 & 8.05-8.9  & 	 8.05-8.9  & 8.05-8.9   \\ 
branch\tablenotemark{c} &    up; low  \phn \phd \phn \phd    & 	       &  up; low  \phn \phd \phn \phd  &  & & & & \\ 
 {\it a } & -461.2352; -4.5710  \phm{} & 	-472.8841 & 	-69.7024; -1.0200  \phm{} & 	1.3782 & 	 & 	193.9000 & 	-738.1193 & 	-1661.9380 \\
 {\it b} & 158.44840; 1.53261 & 	158.20310 & 	16.68313; 1.13063 & 	0.52591 & 	 & 	-64.87895 & 	258.96730 & 	585.17650 \\
 {\it c} & -17.946070; \phm{000000;}  & 	-17.414190 & 	-0.880678; \phm{000000;} & 	0.036003 & 	 & 	7.411102 & 	-30.057050 & 	-68.471750 \\
 {\it d} & 0.6828170; \phm{000000;}  & 	0.6427315 & 	;  \phm{000000;} & 	 & 	 & 	-0.2756653 & 	1.1679370 & 	2.6766690 \\
 $\rho_r$ & 0.060; 0.060  \phm{0;} & 	0.062 & 	0.062; 0.063  \phm{0;} & 	0.051 & 	 & 	0.072 & 	0.058 & 	0.072 \\ \hline
D02 &  &  &  & & & & 	&  \\
x-range\tablenotemark{b} & 8.2-9.2 & 	 8.4-9.3  & 	 8.5-9.1 ;  8.0-8.4   & 8.05-9.2  & 8.05-9.2  &  & 8.05-8.9  & 	 8.05-8.9  \\ 
branch\tablenotemark{c} &  & 	& 	 up; low   \phn \phd \phn \phd &   & &  &   & 	  \\ 
 {\it a} & 1202.5280 & 	-114.3143 & 	-121.3958; -1.0361  \phm{} & 	680.5167 & 	253.0031 & 	 & 	-1.3992 & 	-629.0499 \\
 {\it b} & -412.30820 & 	32.79523 & 	28.86410; 1.13468 & 	-235.27350 & 	-87.03697 & 	 & 	-5.32702 & 	215.37940 \\
 {\it c} & 47.332730 & 	-2.782591 & 	-1.599613; \phm{000000;} & 	27.345830 & 	10.241880 & 	 & 	1.562757 & 	-24.305910 \\
 {\it d} & -1.8065160 & 	0.0731175 & 	;  \phm{000000;} & 	-1.0552390 & 	-0.3984731 & 	 & 	-0.0938063 & 	0.9168766 \\
 $\rho_r$ & 0.058 & 	0.056 & 	0.061; 0.093  \phm{0;} & 	0.038 & 	0.044 & 	 & 	0.033 & 	0.000 \\ \hline
PP04\, (O3N2) &  &  &  & & & & 	&  \\
x-range\tablenotemark{b} & 8.2-9.2 & 8.4-9.3  & 8.5-9.1 ;  8.05-8.4   & 8.1-9.2  & 8.05-9.2  & 8.05-8.9  & 	  & 	 8.05-8.8   \\ 
branch\tablenotemark{c} &  &  & up; low   \phn \phd \phn \phd &   &   &   & 	  & 	   \\ 
 {\it a} & 631.2766 & 	52.2389 & 	-65.0991; 2.1063  \phm{} & 	664.8453 & 	230.7820 & 	36.6598 & 	 & 	-8.0069 \\
 {\it b} & -210.02090 & 	-18.67559 & 	15.74995; 0.74427 & 	-225.75330 & 	-75.79752 & 	-7.64786 & 	 & 	2.74353 \\
 {\it c} & 23.483050 & 	2.447698 & 	-0.837514; \phm{000000;} & 	25.768880 & 	8.526986 & 	0.508480 & 	 & 	-0.093680 \\
 {\it d} & -0.8704286 & 	-0.1011578 & 	;  \phm{000000;} & 	-0.9761368 & 	-0.3162894 & 	 & 	 & 	 \\
 $\rho_r$ & 0.050 & 	0.047 & 	0.056; 0.073   \phm{0;} & 	0.047 & 	0.046 & 	0.045 & 	 & 	0.038 \\ \hline
PP04\, (N2) &  &  &  & & & & 	&  \\
x-range\tablenotemark{b} & 8.2-9.2 & 8.4-9.3  & 8.5-9.1 ;  8.05-8.4  & 8.05-9.2  & 8.05-9.2  & 8.05-8.9  & 8.05-8.9  & 	   \\ 
branch\tablenotemark{c} & &   & up; low   \phn \phd \phn \phd &  & &   &   & 	   \\ 
 {\it a} & 916.7484 & 	656.5128 & 	1334.9130; 3.1447  \phm{} & 	569.4927 & 	319.7570 & 	-444.7831 & 	512.7575 & 	 \\
 {\it b} & -309.54480 & 	-224.11240 & 	-464.86390; 0.61788 & 	-192.51820 & 	-107.13160 & 	165.42600 & 	-180.47540 & 	 \\
 {\it c} & 35.051680 & 	25.734220 & 	54.166750; \phm{000000;} & 	21.918360 & 	12.208670 & 	-20.202000 & 	21.415880 & 	 \\
 {\it d} & -1.3188000 & 	-0.9812624 & 	-2.0986640;  \phm{000000;} & 	-0.8278840 & 	-0.4606539 & 	0.8249386 & 	-0.8427312 & 	 \\
 $\rho_r$ & 0.068 & 	0.065 & 	0.071; 0.050   \phm{0;} & 	0.045 & 	0.051 & 	8.557e-6 & 	0.042 & 	 \\ \hline
\enddata
\tablenotetext{a}{The conversion constants convert metallicities from the calibrations given in row 1 ($x$)
into a metallicity that is consistent with the calibration given in column 1 ($y$), using $y=a+bx+cx^2 + dx^3$.}
\tablenotetext{b}{The range over which a metallicity using a calibration given in row 1 ($x$) can be converted into a metallicity that is consistent with the calibration given in column 1 ($y$).}
\tablenotetext{c}{Some conversions are different, depending on whether the input metallicity ($x$) is on the upper or lower branch (up; low). In some regimes, both the upper and lower branch conversion equations are valid.  In these cases, additional information such an \NIIOII\ or \NIIHa\ ratio is needed to distinguish between the upper and lower \R23\ branches.  A worked example for this case is given in Appendix~\ref{worked_examples}}
\end{deluxetable}
\clearpage
\end{landscape}

\pagestyle{empty}

\begin{deluxetable}{ll}
\tablewidth{0pt}
\tablecolumns{2}
\tabletypesize{\scriptsize}
\tablecaption{Residual Metallicity Discrepancy \label{dev_table}}
\tablehead{Base & Mean Residual\\
ID\tablenotemark{a} & Discrepancy\tablenotemark{b} (dex)}
\startdata
 KK04 & 0.011  \\
 Z94 &  0.018 \\
 M91 & 0.011 \\
 KD02 & 0.014  \\
 T04 & 0.034 \\
 D02 & 0.034 \\
 PP04 O3N2 & 0.012 \\
 PP04 N2  & 0.021  
\enddata
\tablenotetext{a}{calibration into which the other 7 MZ relations have been converted}
\tablenotetext{b}{Mean residual metallicity discrepancy between the 7 converted MZ relations and the base MZ relation.}
\end{deluxetable}
\clearpage

\begin{landscape}
\begin{deluxetable}{lllllllllll}
\tablecolumns{10}
\tablewidth{0cm}
\tabletypesize{\scriptsize}
\tablecaption{Relative median metallicity and rms scatter for the 9 strong-line metallicity calibrations from 30,000 random sets of SDSS galaxies \label{relative_rms_table}}
\tablehead{ ID &  T04 & Z94 & KK04 & KD02 & M91 & D02 & PP04 (O3N2) & PP04 (N2) & P01 & Mean \\  & med (rms) & med (rms) & med (rms) & med (rms) & med (rms) & med (rms) & med (rms) & med (rms) & med (rms) & rms \\}
\startdata
T04 & \nodata & 0.006 (0.08) & 0.053 (0.09) & 0.033 (0.07) & 0.036 (0.09) & 0.076 (0.10) & 0.040 (0.08) & 0.053 (0.09) & 0.046 (0.14) & 0.09 \\
Z94 & -0.006 (0.08) &  \nodata & 0.046 (0.05) & 0.023 (0.05) & 0.024 (0.04) & 0.062 (0.09) & 0.029 (0.06) & 0.042 (0.08) & 0.034 (0.11)  & 0.07\\
KK04 & -0.053 (0.09) & -0.046 (0.05) &  \nodata & -0.023 (0.05) & -0.014 (0.02) & 0.016 (0.06) & -0.017 (0.05) & -0.003 (0.07) & -0.004 (0.09) & 0.06 \\
KD02 & -0.033 (0.07) & -0.023 (0.05) & 0.023 (0.05) &  \nodata & 0.005 (0.05) & 0.039 (0.06) & 0.006 (0.05) & 0.019 (0.05) & 0.011 (0.10) & 0.06 \\
M91 & -0.036 (0.09) & -0.024 (0.04) & 0.014 (0.02) & -0.005 (0.05) &  \nodata & 0.031 (0.08) & -0.001 (0.06) & 0.009 (0.07) & 0.010 (0.08) & 0.06 \\
D02 & -0.076 (0.10) & -0.062 (0.09) & -0.016 (0.06) & -0.039 (0.06) & -0.031 (0.08) &  \nodata & -0.033 (0.05) & -0.020 (0.03) & -0.017 (0.11) & 0.07 \\
PP04 (O3N2) & -0.040 (0.08) & -0.029 (0.06) & 0.017 (0.05) & -0.006 (0.05) & 0.001 (0.06) & 0.033 (0.05) &  \nodata & 0.014 (0.05) & 0.012 (0.12) & 0.07 \\
PP04 (N2) & -0.053 (0.09) & -0.042 (0.08) & 0.003 (0.07) & -0.019 (0.05) & -0.009 (0.07) & 0.020 (0.03) & -0.014 (0.05) &  \nodata & 0.004 (0.11) & 0.07 \\
P01 & -0.046 (0.14) & -0.034 (0.11) & 0.004 (0.09) & -0.011 (0.10) & -0.010 (0.08) & 0.017 (0.11) & -0.012 (0.12) & -0.004 (0.11) &  \nodata & 0.11\\
\enddata
\end{deluxetable}
\clearpage
\end{landscape}

\end{document}